\documentclass[12pt,preprint]{aastex}

\newcommand{\CIVdblt}{{C}\kern 0.1em{\sc iv}~$\lambda\lambda 1548, 1550$}
\newcommand{\MgIIdblt}{{Mg}\kern 0.1em{\sc ii}~$\lambda\lambda 2796, 2803$}
\newcommand{\SiIVdblt}{{Si}\kern 0.1em{\sc iv}~$\lambda\lambda 1393, 1402$}
\newcommand{\NVdblt}{\hbox{{N}\kern 0.1em{\sc v}~$\lambda\lambda 1239,1243$}}
\newcommand{\OVIdblt}{{O}\kern 0.1em{\sc vi}~$\lambda\lambda 1032, 1038$}
\newcommand{\AlIIIdblt}{{Al}\kern 0.1em{\sc iii}~$\lambda\lambda 1855, 1863$}
\newcommand{\CII}{\hbox{{C}\kern 0.1em{\sc ii}}}
\newcommand{\CIII}{\hbox{{C}\kern 0.1em{\sc iii}}}
\newcommand{\CIV}{\hbox{{C}\kern 0.1em{\sc iv}}}
\newcommand{\HI}{\hbox{{H}\kern 0.1em{\sc i}}}
\newcommand{\NaI}{\hbox{{Na}\kern 0.1em{\sc i}}}
\newcommand{\HII}{\hbox{{H}\kern 0.1em{\sc ii}}}
\newcommand{\HeI}{\hbox{{He}\kern 0.1em{\sc i}}}
\newcommand{\HeII}{\hbox{{He}\kern 0.1em{\sc ii}}}
\newcommand{\AlII}{\hbox{{Al}\kern 0.1em{\sc ii}}}
\newcommand{\AlIII}{\hbox{{Al}\kern 0.1em{\sc iii}}}
\newcommand{\NII}{\hbox{{N}\kern 0.1em{\sc ii}}}
\newcommand{\Lya}{\hbox{{Ly}\kern 0.1em$\alpha$}}
\newcommand{\Lyb}{\hbox{{Ly}\kern 0.1em$\beta$}}
\newcommand{\Lyg}{\hbox{{Ly}\kern 0.1em$\gamma$}}
\newcommand{\Lyd}{\hbox{{Ly}\kern 0.1em$\delta$}}
\newcommand{\Lye}{\hbox{{Ly}\kern 0.1em$\epsilon$}}
\newcommand{\FeII}{\hbox{{Fe}\kern 0.1em{\sc ii}}}
\newcommand{\MgI}{\hbox{{Mg}\kern 0.1em{\sc i}}}
\newcommand{\MgII}{\hbox{{Mg}\kern 0.1em{\sc ii}}}
\newcommand{\OI}{\hbox{{O}\kern 0.1em{\sc i}}}
\newcommand{\OVI}{\hbox{{O}\kern 0.1em{\sc vi}}}
\newcommand{\OVII}{\hbox{{O}\kern 0.1em{\sc vii}}}
\newcommand{\OVIII}{\hbox{{O}\kern 0.1em{\sc viii}}}
\newcommand{\NV}{\hbox{{N}\kern 0.1em{\sc v}}}
\newcommand{\SiII}{\hbox{{Si}\kern 0.1em{\sc ii}}}
\newcommand{\SiIII}{\hbox{{Si}\kern 0.1em{\sc iii}}}
\newcommand{\SiIV}{\hbox{{Si}\kern 0.1em{\sc iv}}}
\newcommand{\kms}{\ensuremath{\mathrm{km~s^{-1}}}}
\newcommand{\cmsq}{\ensuremath{\mathrm{cm^{-2}}}}
\newcommand{\cc}{\ensuremath{\mathrm{cm^{-3}}}}
\newcommand{\Ang}{\hbox{\textrm{\AA}}}

\newcommand{\Zsun}{\ensuremath{Z_{\Sun}}}

\begin{document}

\title{Physical Properties of Weak \MgII\ Absorbers at $z \sim 2$\footnotemark[1]}

\footnotetext[1]{Based on public data obtained from the ESO archive of observations from the UVES spectrograph at the VLT, Paranal, Chile.  ESO Program ID 166.A-0106.  HE2217-2818 observed during UVES commissioning.}

\author{Ryan S. Lynch}
\affil{Department of Astronomy, P.O Box 400325, University of Virginia, Charlottesville, VA 22904}
\email{rsl4v@virginia.edu}

\and

\author{Jane C. Charlton}
\affil{Department of Astronomy and Astrophysics, The Pennsylvania State University, University Park, PA 16802}
\email{charlton@astro.psu.edu}




\begin{abstract}
We present the results of photoionization modeling of nine weak \MgII\ ($W_r < 0.3~\Ang$) quasar absorption line systems with redshifts $1.4 < z < 2.4$ obtained with the  Ultraviolet and Visual Echelle Spectrograph on the Very Large Telescope.  These systems have been chosen because they provide access to a regime of red-shift space that previous weak \MgII\ studies have not looked at.  The densities, metallicities, Doppler parameters, and column densities of these systems are compared to those of other weak \MgII\ systems at lower redshift.  There is no significant statistical variation in the properties of the absorbers over the redshift range $0.4 < z < 2.4$.  The number density per unit redshift is known to decrease for weak \MgII\ absorbers between $z \sim 1$ and $z \sim 2$ by a greater amount than predicted from cosmological effects and changes in the extragalactic ionizing background alone.  We suggest that, because the physical properties of the absorber population are not seen to change significantly across this range, that the evolution in $dN/dz$ is due to a decrease in the activity that gives rise to weak \MgII\ absorption, and not due to a change in the processes that form weak \MgII\ absorbers.  The presence of separate, but aligned (in velocity) low and high density clouds in all single cloud weak \MgII\ absorbers provides an important diagnostic of their geometry.  We discuss possible origins in dwarf galaxies and in extragalactic analogs to high velocity clouds.
\end{abstract}


\section{Introduction}
\label{sec:intro}

Weak \MgII\ absorbers are defined to be those with rest frame equivalent widths $W_r < 0.3~\Ang$ and represent a different population than strong \MgII\ absorbers \citep{Rig02, Nest05}.  Strong \MgII\ absorbers are known to be associated with luminous galaxies (within $\sim 38 h^{-1} (L/L^*)^{0.15}$ kpc) \citep{Berg91, Berg92, LeBrun93, Steid94, Steid97, Steid95}, while weak \MgII\ absorbers are not typically seen within a $50 h^{-1}$ kpc impact parameter of a luminous galaxy \citep[but see \citet{Church05} for some exceptions]{Rig02}.  The exact environment(s) and process/processes that give rise to weak \MgII\ absorbers is not yet known, but they may arise in dwarf galaxy environments, in the cosmic web surrounding galaxies, and/or in high velocity clouds.  Weak \MgII\ absorbers generally correspond to sub-Lyman limit systems ($15.8 < \log{N(\HI)} < 16.8 [\cmsq]$) \citep{Church99b,Rig02,Church00} and they have metallicities of at least 10\% solar and as high as solar or even supersolar \citep{Rig02,Char03,Sim05}.  In addition, the \FeII/\MgII\ ratio of some absorbers does not allow for $\alpha$-enhancement, thus Type Ia supernovae must contribute as well as Type II.  Because Type Ia supernovae cannot eject metals to large distances, metals must be produced ``\textit{in situ}''.

The number statistics and kinematics of single cloud weak \MgII\ absorbers lend themselves best to a flattened geometry and suggest that the absorbers may be produced by higher density regions in the cosmic web \citep{Milni05}.  \MgII\ absorption typically arises in a high density region $\sim$ 1--100 pcs thick which is often surrounded by a lower density region that gives rise to high ionization \CIV\ absorption centered at the same velocity as the \MgII\ \citep{Char03,Sim05}.  Additional low density regions producing \CIV\ absorption, that are detected at different velocities than \MgII, often exist.  

While most weak \MgII\ absorbers can be fit by a single Voigt profile component, about one third have multiple components \citep{Church00,Lynch06}.  Some of these multiple cloud weak \MgII\ absorbers are weaker versions of strong \MgII\ absorbers with similar kinematics, as if they arise in the outskirts or in sparse regions of luminous galaxies.  Others, which tend to have more kinematically compact profiles, have been hypothesized to arise in dwarf galaxies \citep{Zonak04,Ding05,Mas05}.

When exploring the nature of weak \MgII\ absorbers, two methods are often used--statistical surveys and photoionization modeling of individual systems.  \citet{Nar05}, \citet{Church99b}, and \citet{Lynch06} conducted surveys of systems with equivalent widths in the range $0.02 \le W_r \le 0.3~\Ang$, which when combined, span the redshift range $0 < z < 2.4$.  These surveys obtained number densities of absorbers per unit redshift ($dN/dz$): $dN/dz = 1.00 \pm 0.20~(0 < z < 0.3)$ \citep{Nar05}, $dN/dz = 1.74 \pm 0.10~(0.4 < z < 1.4)$ \citep{Church99b}, and $dN/dz = 1.06 \pm 0.12~(1.4 < z < 2.4)$ \citep{Lynch06}.  Apparently, the population of weak \MgII\ absorbers peaks at $z \sim 1$ over the range $0 < z < 2.4$.  Furthermore, the density of absorbers at $z \sim 2$ is significantly lower than expected if the change in $dN/dz$ was due only to cosmological effects and to the changing extragalactic background radiation \citep{Lynch06}.  This indicates that either the same process creates weak \MgII\ absorbers across this redshift range but was less active at $z \sim 2$ than at $z \sim 1$, or that the physical mechanisms responsible for creating the absorbers change across redshift and are more efficient at $z \sim 1$ than at $z \sim 2$.  It should be noted that this trend in $dN/dz$ is consistent with the star formation history in dwarf galaxies \citep{Gab04,Kauffmann04}, which may suggest that weak \MgII\ absorbers are related to this activity.

Photoionization models facilitate the exploration of the physical properties of the absorption system, such as ionization parameter, density, temperature, abundance pattern, and size. By understanding these properties for absorbers at different redshifts, we can gain insight into what processes create them.   If different processes are responsible for the formation of these absorbers at different epochs, then the physical properties of the absorbers are likely to evolve.  If no statistical difference in the physical properties is observed, this would suggest that the same mechanism is responsible for creating weak \MgII\ absorbers at different epochs.  To this end, we have produced photoionization models of the nine $1.4 < z < 2.4$ systems found by \citet{Lynch06} using the code Cloudy \citep{Fer98} and compared our results to those of systems at lower redshift \citep{Rig02,Char03,Zonak04,Ding05,Mas05}.  Our methodology for modeling the systems is the subject of \S~\ref{sec:method}.  In \S~\ref{sec:props} we report the results of our models for each individual system.  In \S~\ref{sec:disc} we discuss our results and their implications for the population of weak \MgII\ absorbers and end with a conclusion in \S~\ref{sec:conc}.

\section{Modeling Methodology}
\label{sec:method}

The 9 systems selected for modeling were taken from \citet{Lynch06}, and were identified through a detection of a weak \MgIIdblt\ doublet in the redshift range $1.4 < z < 2.4$.  The systems were chosen because they lie in a region of redshift-space ($1.4 < z < 2.4$) that previous weak \MgII\ studies have not looked at.  In order to be accepted as a true detection, the \MgIIdblt\ had to be detected at at least a $5\sigma$ significance level in \MgII\ $\lambda$2796.  For a detailed discussion of the data reduction procedure see \citet{Kim04}, and for a discussion of the survey method see \citet{Lynch06}.  Once an absorption system was identified, the Doppler parameter and column density of the \MgIIdblt\ doublet were measured from a Voigt profile fit to the doublet using Minfit \citep{Church03}.  This code finds the minimum number of components required for an adequate fit to the observed absorption of \MgII.  

The measured \MgII\ column density was used as a direct constraint for the photoionization code Cloudy 94.00 last described by \citet{Fer98} (i.e. the photoionization models were ``optimized'' on the column density of \MgII).  For each cloud (separate Voigt profile component) a grid was constructed for which the metallicity ($\log{Z/\Zsun}$) and ionization parameter ($\log{U} = \log{n_e/n_{\gamma}}$) were changed in incremental values, typically in steps of 0.5 dex.  A solar abundance pattern was assumed unless otherwise noted. A Haardt \& Madau ionizing background of quasars plus star forming galaxies with a photon escape fraction of 0.1 for $z<3$, was used at the redshift of each system \citep{Haardt96,Haardt01}. This radiation was assumed to be incident on a plane parallel slab. Cloudy calculated column densities for all detected transitions and an equilibrium temperature for the cloud.  This temperature was used along with the measured Doppler parameter of \MgII\ to calculate the turbulent/bulk motion contribution to the total Doppler parameter, $b_{turb}^2 = b_{tot}^2 - 2kT/m_{Mg}$.  This $b_{turb}$ was then applied to calculate the expected $b_{tot}$ for every other element, again using the equilibrium temperature given by Cloudy.  The model column densities and Doppler parameters of all clouds were used to generate a synthetic spectrum which was convolved with the instrumental profile appropriate for UVES, $R = 45,000$.  These were compared to the data by using $\chi^2$ indicators combined with profile inspection in order to refine the metallicity and ionization parameter values of the model.  At times, it was necessary to adjust the abundance pattern of the model in order to achieve an adequate fit to the observed spectrum.  The effect of such changes on model parameters is given in the individual system descriptions.  We assume a solar abundance pattern unless otherwise stated.

Many of the low wavelength transitions fell in the \Lya\ forest and contained significant blends.  In this case, a model was taken to be acceptable if it did not overproduce the observed absorption of a given transition.  However, when blends were not present, a more exact match was required.  In some cases a range of model parameters produced an adequate fit.  In the case of blends, an upper or lower limit was often the only constraint that could be obtained.

In most cases, there was absorption in higher ionization transitions that Cloudy could not reproduced through the low ionization \MgII\ phase (i.e. the clouds that were optimized on the \MgII\ column density).  This happened either because the high ionization absorption was too broad to arise from the low ionization phase, or because the high ionization absorption was not sufficiently produced for ionization parameters that provided a sufficient fit to low and intermediate ionization absorption.  In these cases, the Doppler parameters and column densities of the \CIVdblt\ doublet were measured using Minfit and Cloudy models were produced, optimizing on \CIV. The combined low and high ionization phase models were compared to the data.   Separate constraints on metallicity and ionization parameter were obtained for the high ionization clouds, when possible.  However, it is common that the low ionization phase has only a lower limit on metallicity in order not to overproduce \Lya\ absorption.  For higher values of metallicity, an additional contribution to \Lya\ from the high ionization phase would be required.  In the event that there was a blend in the expected location of the \CIVdblt\ doublet, the \SiIVdblt\ doublet was used instead.  If there were blends in the expected locations of both these transitions, no constraint on the high ionization phase could be obtained (though this was a problem for only the $z=1.708494$ system toward HE0151-4326).

In addition to a Haardt \& Madau ionizing background of quasars plus star forming galaxies with a photon escape fraction of 0.1 for $z<3$, the effect of using an ionizing background including only quasars was also explored.  In most cases, this change had little effect, and the small effect it did have was only seen in the high ionization gas.  The sole exception is the $z = 1.450109$ absorber towards Q0122-380, for which the effect of the change of spectral shape is described in \S~\ref{sec:S3}.

\section{Properties of Individual Systems}
\label{sec:props}

\begin{figure}
\centering
\epsscale{0.9}
\plotone{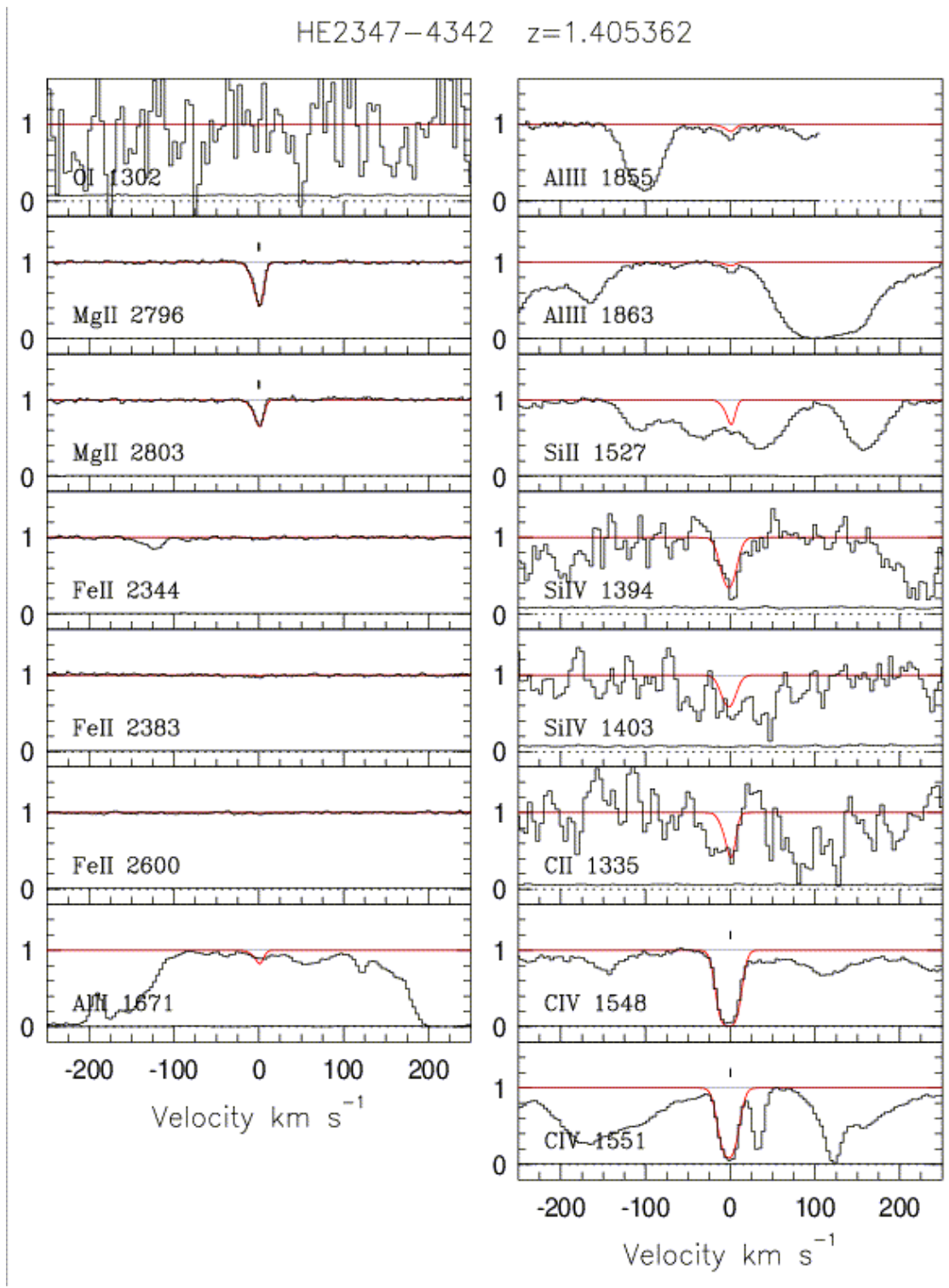}
\caption[A best fit model of the $z = 1.405362$ system towards HE2347-4342.]{\small The $z = 1.405362$ system towards HE2347-4342.  Only relevant transitions are displayed here, in velocity space, centered at the redshift corresponding to the optical depth weighted mean of the \MgII\ $\lambda$2796 profile.  The error array is plotted, but in some cases is so small that it is difficult to distinguish from zero flux.  This model uses $\log{U} = -3.0$ for the low ionization phase and $\log{U} = -1.7$ for the high ionization phase, and $\log{Z/\Zsun} = 0.0$ for both phases. The positions of model clouds from the low ionization phase are marked with ticks on the \MgII\ panels, while the high ionization phase clouds are marked with ticks on the \CIV\ panels.  The feature to the right of \CIV\ $\lambda$1551 is most likely \Lya\ absorption in the forest.  \label{fig:S1}}
\end{figure}

\begin{figure}
\centering
\epsscale{1.0}
\plotone{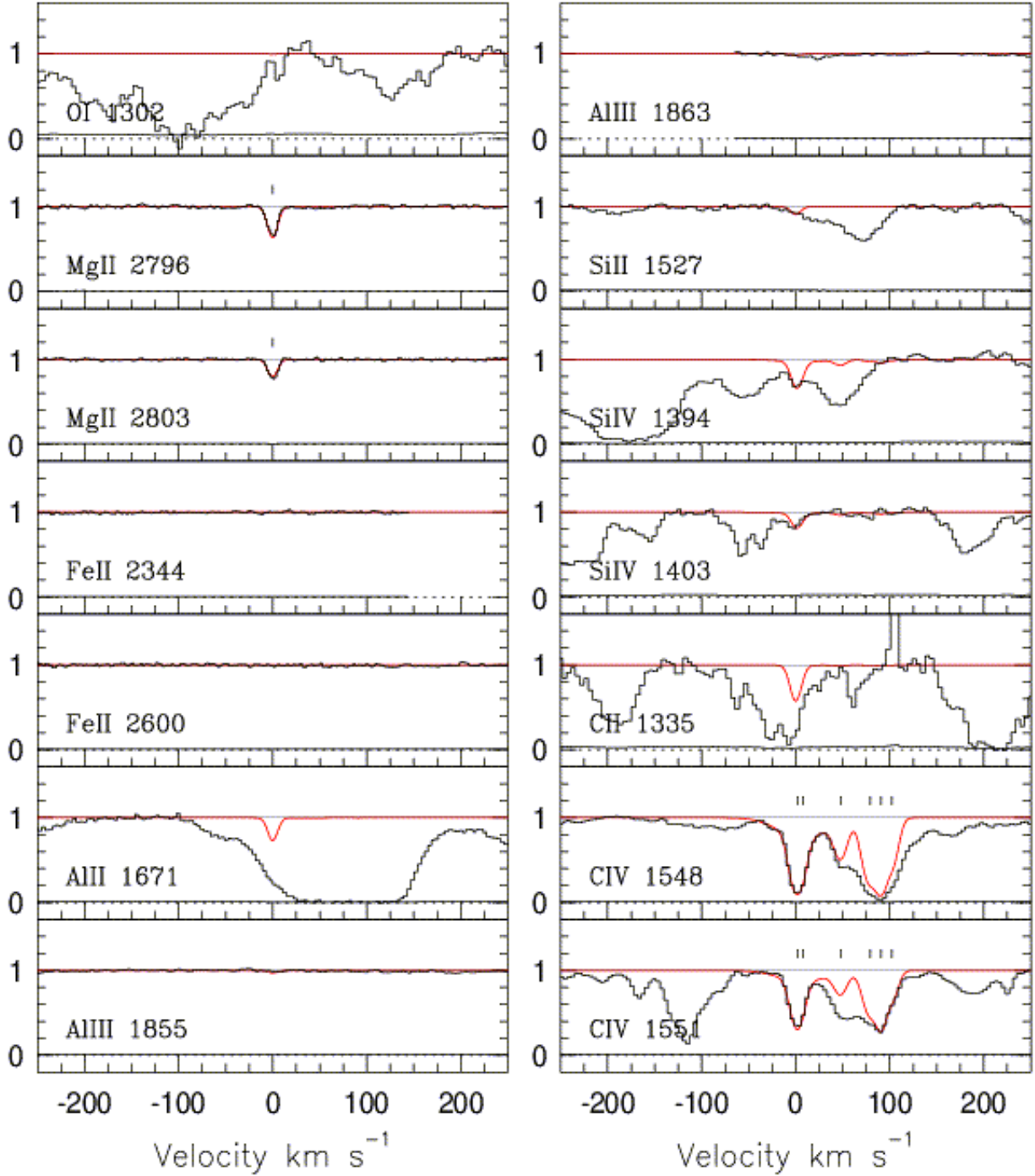}
\caption[A best fit model of the $z = 1.446496$ system towards Q0002-422.]{\small The $z = 1.446496$ system towards Q0002-422, displayed as in Fig.~\ref{fig:S1}.  Only relevant transitions are displayed here.  This model uses $\log{U} = -3.0$ for the low ionization phase and $\log{U} = -1.3$ for the high ionization phase.  Metallicity is $\log{Z/\Zsun} = -1.0$ \label{fig:S2}}
\end{figure}

\begin{figure}
\centering
\epsscale{1.0}
\plotone{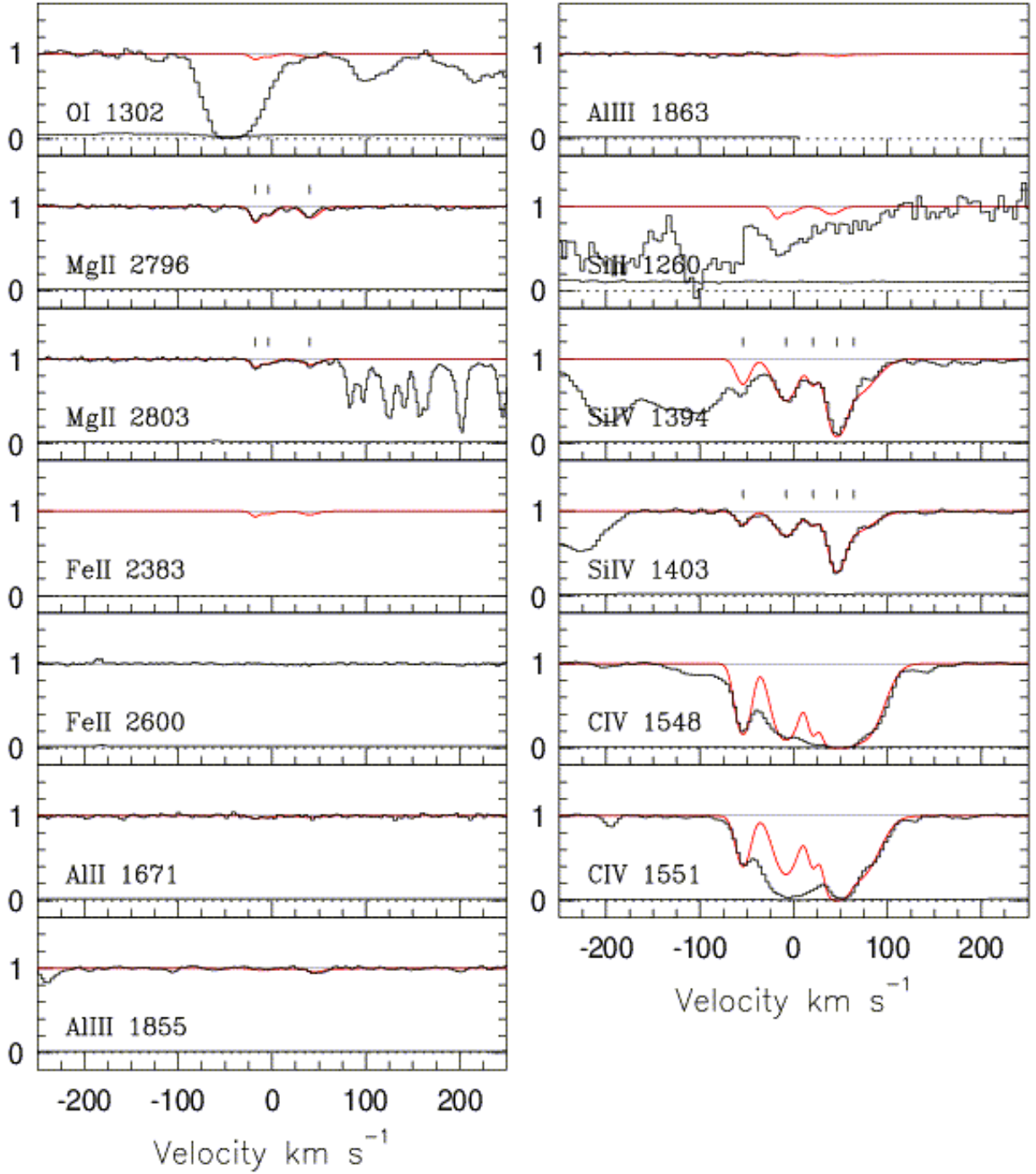}
\caption[A best fit model of the $z = 1.450109$ system towards Q0122-380.]{\small The $z = 1.450109$ system towards Q0122-380, displayed as in Fig.~\ref{fig:S1}.  Only relevant transitions are displayed here.  This model uses $\log{U} = -5.0$ for the low ionization phase and $\log{U} = -1.1$ for the high ionization phase.  Metallicity is $\log{Z/\Zsun} = 0.0$ for the low ionization phase and $\log{Z/\Zsun} = 1.0$ for the high ionization phase. \label{fig:S3}}
\end{figure}

\begin{figure}
\centering
\epsscale{1.0}
\plotone{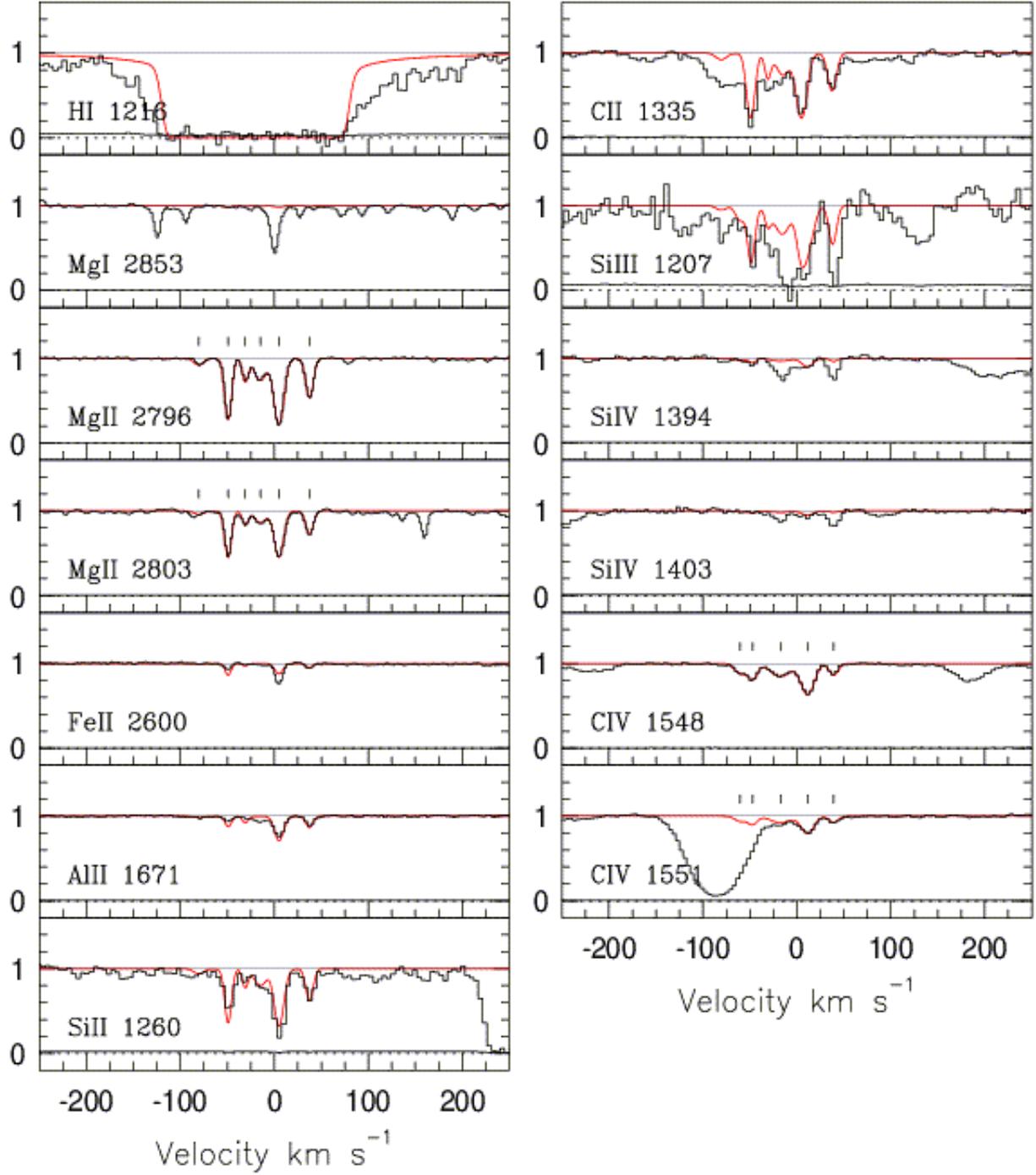}
\caption[A best fit model of the $z = 1.555845$ system towards HE2217-2818.]{\small The $z = 1.555845$ system towards HE2217-2818, displayed as in Fig.~\ref{fig:S1}.  Only relevant transitions are displayed here.  This model uses $\log{U} = -4.0$ for the low ionization phase and $\log{U} = -2.0$ for the high ionization phase.  Metallicity $\log{Z/\Zsun} = 1.4$ \label{fig:S4}}
\end{figure}

\begin{figure}
\centering
\epsscale{1.0}
\plotone{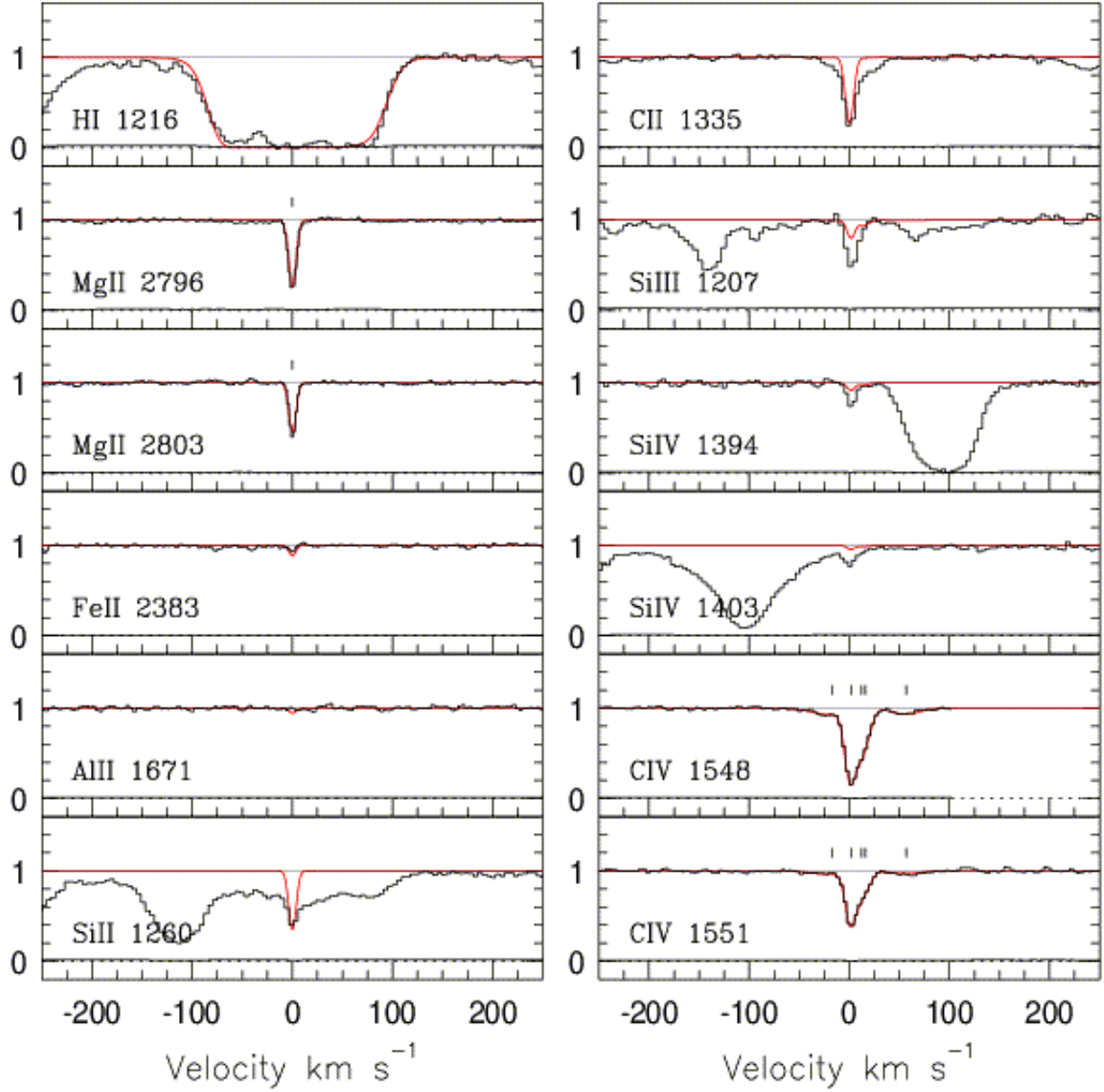}
\caption[A best fit model of the $z = 1.651462$ system towards HE0001-2340.]{\small The $z = 1.651462$ system towards HE0001-2340, displayed as in Fig.~\ref{fig:S1}.  Only relevant transitions are displayed here.  This model uses $\log{U} = -5.0$ for the low ionization phase and $\log{U} = 0.0$ for the high ionization phase.  Metallicity is $\log{Z/\Zsun} = -1.5$ The abundances of Fe, Al, and C have been lowered relative to solar.  \label{fig:S5}}
\end{figure}

\begin{figure}
\centering
\epsscale{1.0}
\plotone{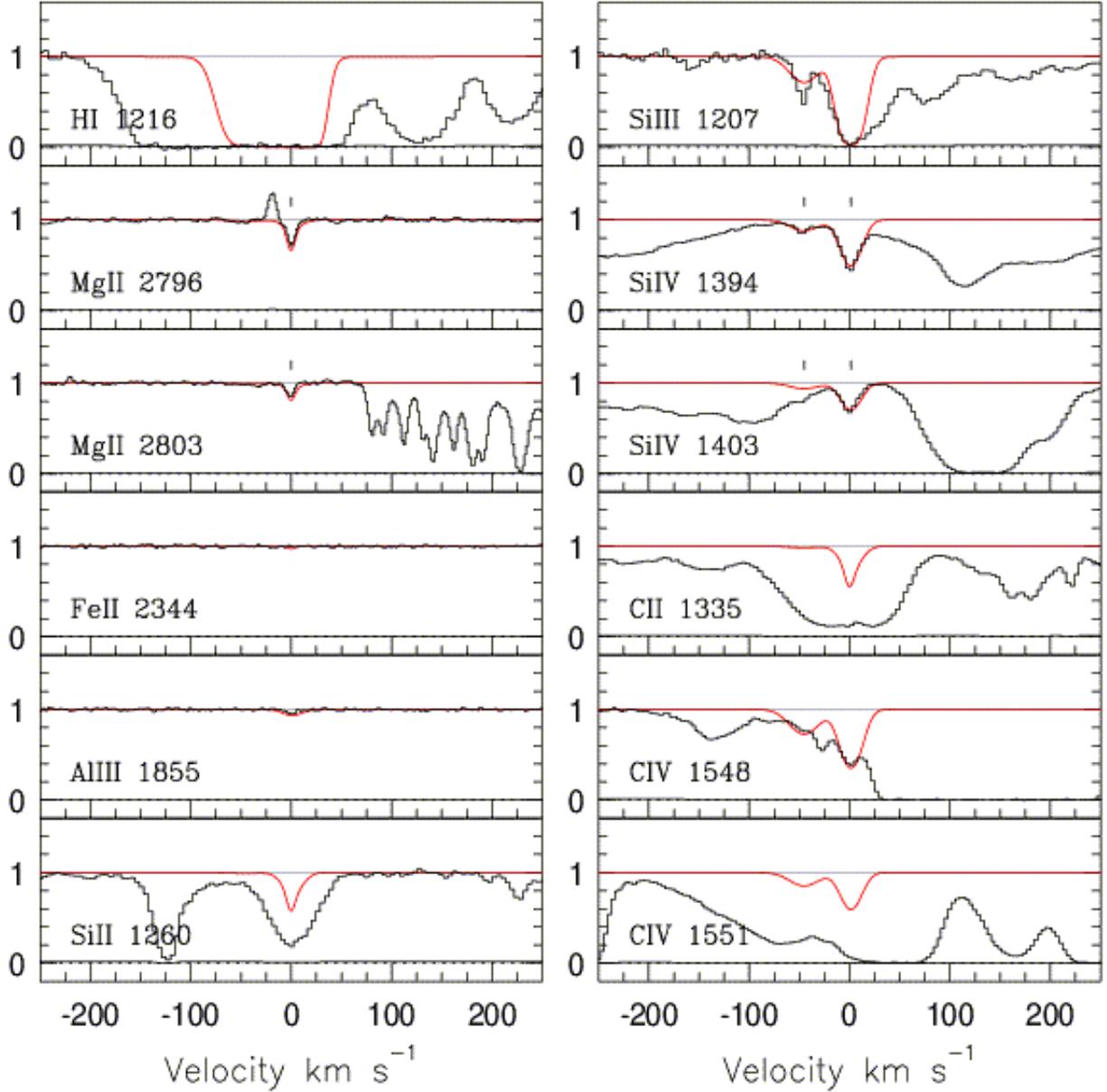}
\caption[A best fit model of the $z = 1.708494$ system towards HE0151-4326.]{\small The $z = 1.708494$ system towards HE0151-4326, displayed as in Fig.~\ref{fig:S1}.  Only relevant transitions are displayed here.  This model uses $\log{U} = -4.0$ for the low ionization phase.  The high ionization phase is not required but here we use it with $\log{U} = -2.0$.  Metallicity is $\log{Z/\Zsun} = -1.0$ The feature to the left of \MgII\ $\lambda$2796 is an artifact of poor sky subtraction.  It has been accounted for in our measurements.  \label{fig:S6}}
\end{figure}

\begin{figure}
\centering
\epsscale{0.9}
\plotone{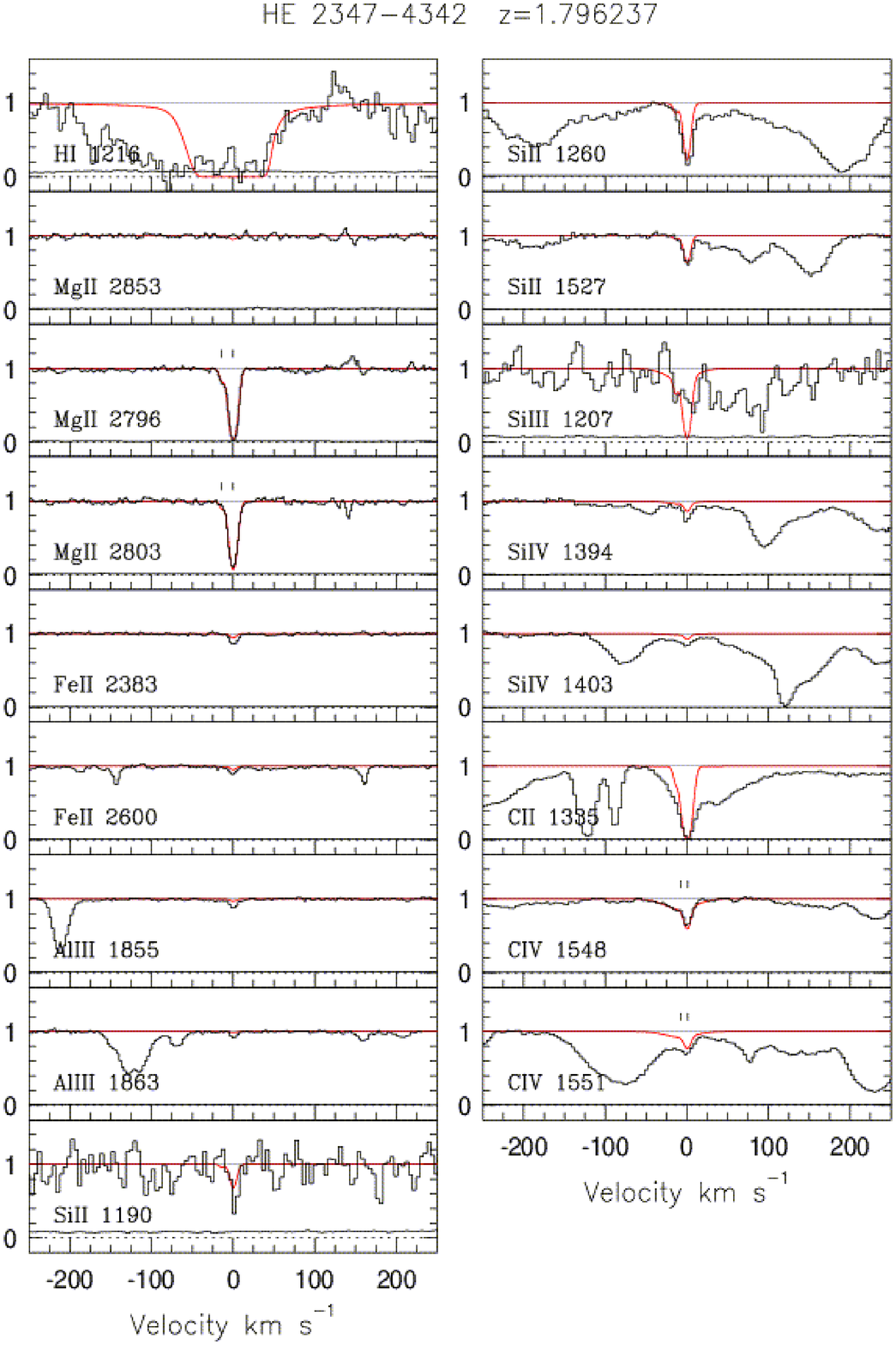}
\caption[A best fit model of the $z = 1.796237$ system towards HE2347-4342.]{\small The $z =1.796237$ system towards HE2347-4342, displayed as in Fig.~\ref{fig:S1}.  Only relevant transitions are displayed here.  This model uses $\log{U} = -4.0$ for the low ionization phase and $\log{U} = -2.0$ for the high ionization phase.  As explained in \S\ref{sec:S7}, a second phase is not definitely needed, though we do use one here.  Metallicity is $\log{Z/\Zsun} = -1.0$ The abundances of Si, Al, and Fe have been lowered relative to solar.  \label{fig:S7}}
\end{figure}

\begin{figure}
\centering
\epsscale{1.0}
\plotone{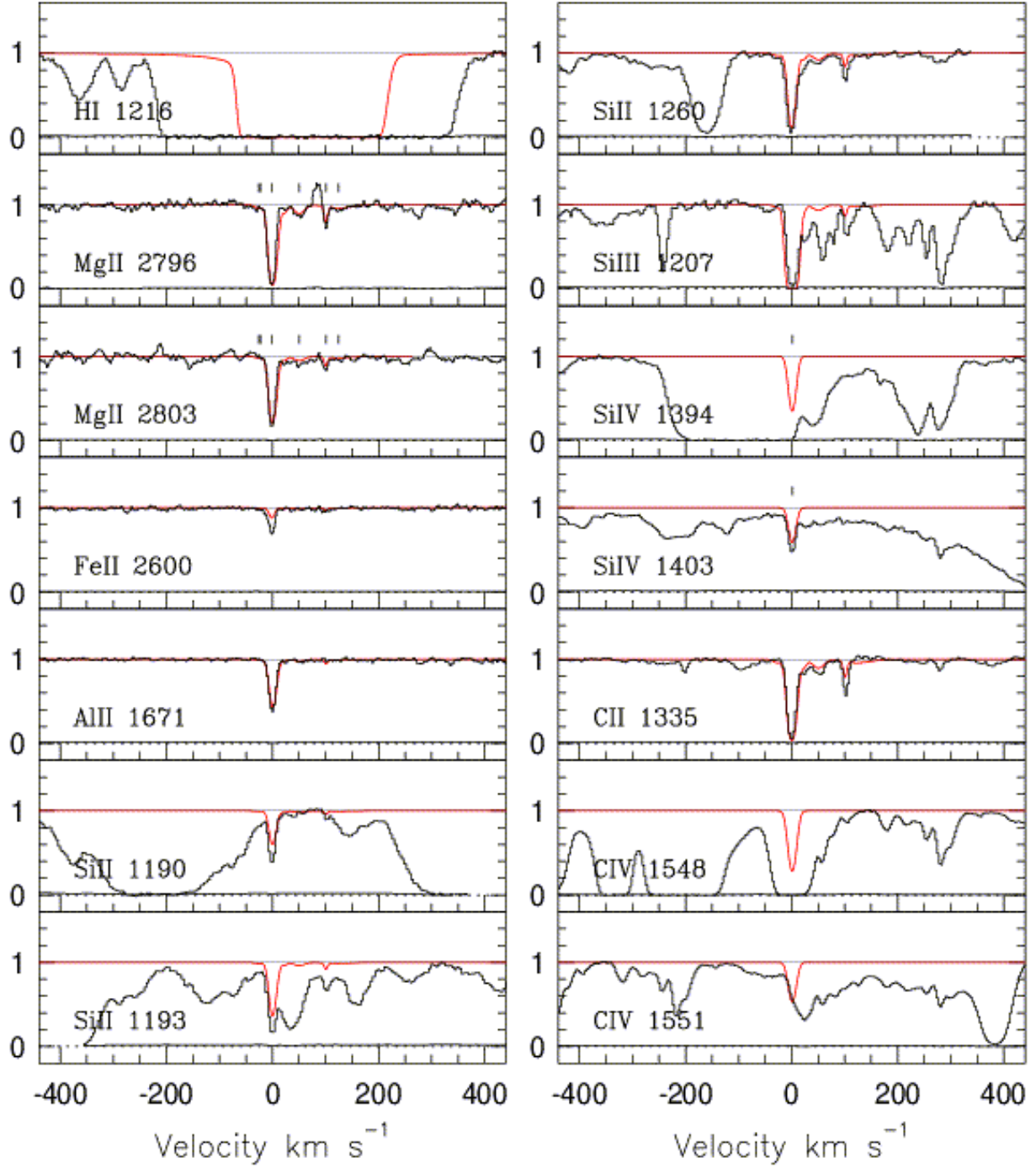}
\caption[A best fit model of the $z = 1.858380$ system towards Q0453-423.]{\small The $z = 1.858380$ system towards Q0453-423, displayed as in Fig.~\ref{fig:S1}.  Only relevant transitions are displayed here.  This model uses $\log{U} = -4.0$ for the low ionization phase.  Here we also use a high ionization phase with $\log{U} = -2.8$.  Metallicity is $\log{Z/\Zsun} = -1.3$ \label{fig:S8}}
\end{figure}

\begin{figure}
\centering
\epsscale{1.0}
\plotone{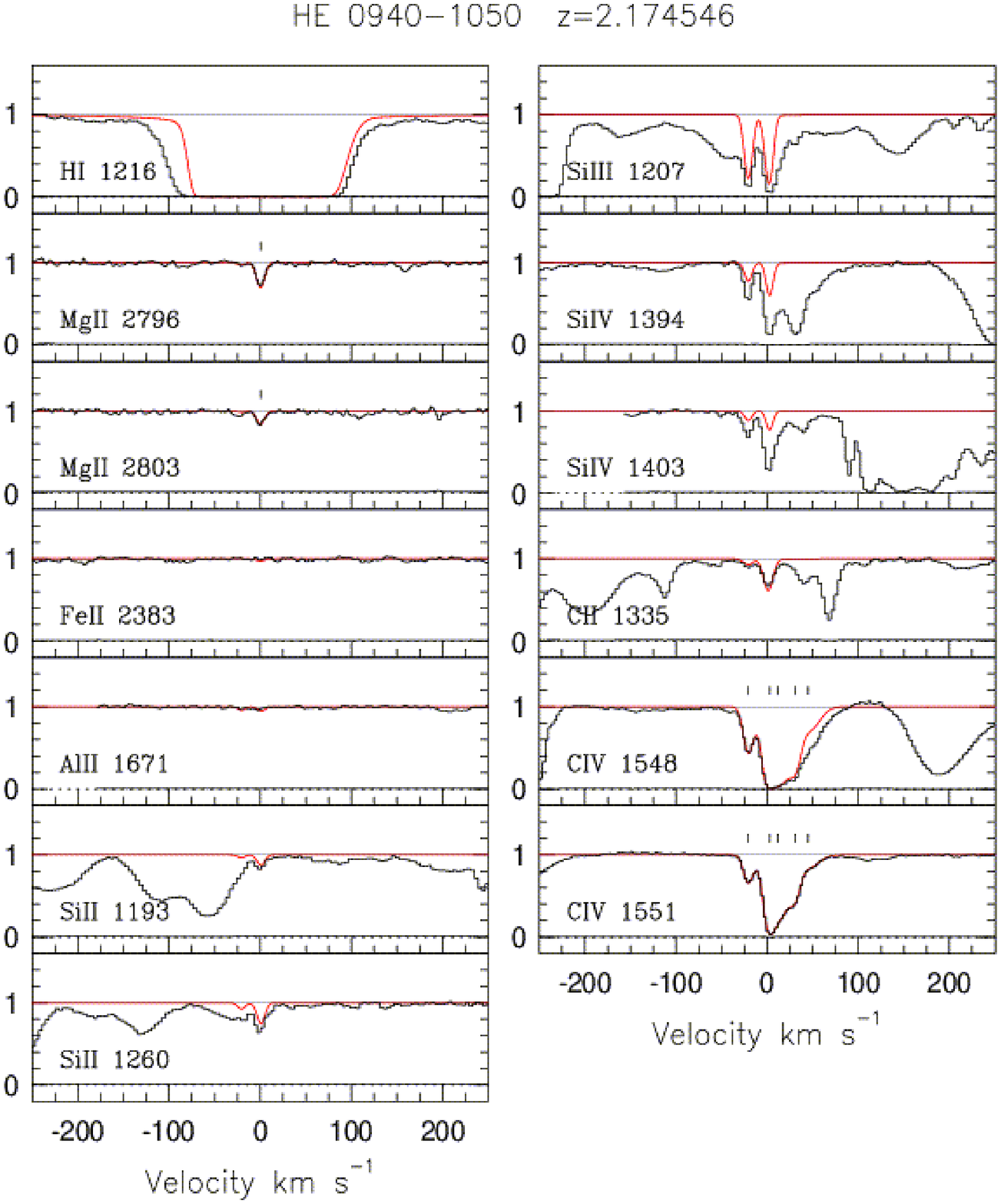}
\caption[A best fit model of the $z = 2.174546$ system towards HE0940-1050.]{\small The $z = 2.174546$ system towards HE0940-1050, displayed as in Fig.~\ref{fig:S1}.  Only relevant transitions are displayed here.  This model uses $\log{U} = -3.7$ for the low ionization phase and $\log{U} = -2.0$ for the high ionization phase.  Metallicity is $\log{Z/\Zsun} = -2.0$ \label{fig:S9}}
\end{figure}

The results of the photoionization modeling of our nine systems are presented here and a summary can be found in Table \ref{table:UandZ}.  Voigt profile fit results with errors obtained using Minfit are presented in Table \ref{table:bandN}.  The best-fit models are superimposed on key constraint transitions in Figures \ref{fig:S1}--\ref{fig:S9}.

\subsection{HE2347-4342 $z=1.405362$}
\label{sec:S1}

This is a single cloud weak \MgII\ absorber, i.e. there is only one resolved component of absorption in \MgII.  A second phase is required to reproduce the observed \CIV\ absorption.  The ionization parameter is constrained to be $-3.0 \le \log{U} \le -2.5$ for the \MgII\ phase and $\log{U} \ge -1.8$ for the \CIV\ phase.  Higher values for the ionization parameter in the low phase give rise to an overproduction of \SiIV\ and \AlIII, and lower values fail to produce enough \AlIII.  A lower ionization parameter in the high phase overproduces \SiIV.  Using a number density of photons of $\log{n_{\gamma}} = -4.83$ at this redshift \citep{Haardt01}, this corresponds to $-2.3 \le \log{n_H} \le -1.8~[\cc]$ for the \MgII\ phase and $\log{n_H} \le -3.1~[\cc]$ for the \CIV\ phase.  This constraint is based upon a solar ratio of Fe to Mg.  However, higher densities would apply if there is $\alpha$-enhancement (i.e. if the Fe to Mg ratio is lower than the solar value).  \Lya\ is not covered due to the low redshift of this system, so there is no constraint on metallicity.  In the optically thin regime, our constraints on $\log U$ are insensitive to the assumed metallicity.

\subsection{Q0002-422 $z=1.446496$}
\label{sec:S2}

This is a single cloud absorber in \MgII. A second phase is required to reproduce the observed \CIV\ absorption.  The ionization parameter is constrained to be $-3.5 \le \log{U} \le -3.0$ for the low ionization phase, and $-1.8 \le \log{U} \le 1.3$ for the high ionization phase.  Lower ionization parameters for the low phase overproduce \OI, and higher values overproduce \SiIV.  The constraint on the high phase is based upon the strength of the \SiIVdblt\ doublet.  Using a $\log{n_{\gamma}} = -4.82$, we find $-1.8 \le \log{n_H} \le -1.3~[\cc]$ for the \MgII\ phase and $-3.5 \le \log{n_H} < -3.0~[\cc]$ for the \CIV\ phase.  For an $\alpha$-enhanced model, lower densities would apply. \Lya\ is not covered due to the low redshift of this system, so there is no constraint on metallicity.

\subsection{Q0122-380 $z=1.450109$}
\label{sec:S3}

This is a multiple cloud absorber in \MgII\ with three resolved components, and requires a second phase to reproduce the observed \CIV\ absorption.  The ionization parameter is constrained to be $-4.0 \le \log{U} \le -2.5$ for the \MgII\ phase across all clouds, and $-1.8 \le \log{U} \le -1.7$ for the \CIV\ phase. Higher values of the ionization parameter for the low phase overproduce \AlIII, and lower values overproduce \OI.  A different ionization parameter for the high phase does not reproduce the observed \SiIV\ absorption.  Using a number density of photons of $\log n_{\gamma} = 4.82$, we find $-2.3 \le \log{n_H} \le -0.8~[\cc]$ for the \MgII\ phase and $-3.2 \le \log{n_H} \le -3.0$ for the \CIV\ phase.  \Lya\ is not covered due to the low redshift of this system, so there is no constraint on metallicity.

For this system, a change of the ionizing spectrum from ``quasar plus star-forming galaxies'' to ``quasar-only''
(see \S~\ref{sec:method}) did have a non-negligible effect.  In the ``quasar-only'' case, the ionization parameter in the high ionization phase needed to be increased by 0.3 dex to $\log{U}\sim-1.5$, and the abundance of aluminum needed to be decreased by 0.5 dex relative to solar.

\subsection{HE2217-2818 $z=1.555845$}
\label{sec:S4}

This is a multiple cloud absorber in \MgII\ with six resolved components, and requires a second phase to reproduce the observed \CIV\ absorption.  The ionization parameter is constrained to be $\log{U} = -4.0$ for the \MgII\ phase, and $-2.5 \le \log{U} \le -2.0$ in the \CIV\ phase. A high ionization parameter for the low phase overproduces \AlIII\ while a lower ionization parameter under-produces the observed \FeII\ absorption.  Other values of the ionization parameter for the high phase cannot reproduce the observed \SiIV\ absorption.  Using a number density of photons of $\log n_{\gamma} = -4.78$, we find $\log{n_H} \le -0.8~[\cc]$ for the \MgII\ phase and $-2.8 \le \log{n_H} \le -2.3~[\cc]$ for the \CIV\ phase.  It is likely that $\alpha$-enhancement is required to explain the observed absorption of Fe and Al  within this range of ionization parameters.  The strength of the \Lya\ line, relative to the low ionization transitions, implies a metallicity of $\log{Z/\Zsun} \ge -1.4$.

\subsection{HE0001-2340 $z=1.651462$}
\label{sec:S5}


This is a narrow, single cloud absorber in \MgII. Two phases are needed to reproduce the observed \CIV\ absorption.  There are two possibilities for fitting the low ionization phase.  One is to use $\log{U} \ge -3.5$ and to decrease the abundances of Si, Al, and C relative to the solar value.  The other possibility is to use $\log{U} \le -4.0$ and to decrease the abundances of Fe, Al, and C relative to solar.  The magnitude of the required adjustment is only about 0.7 dex for each element.  Using a number density of photons of $\log{n_{\gamma}} = -4.8$, we find $\log{n_H} \le -1.3~[\cc]$ and $\log{n_H} \ge -0.8~[\cc]$, respectively, for the two values of ionization parameter.  The higher ionization parameter decreases the observed Fe absorption but overproduces the Si absorption, while the lower ionization parameter decreases the observed Si absorption but overproduces the Fe absorption.  We remark that these abundance patterns are consistent with dust depletion and that this is one possible explanation for the observed absorption profiles \citep{Welt02}.  The ionization parameter is constrained to be $-1.7 \le \log{U} \le -1.5$ for the \CIV\ phase based upon the observed \SiIVdblt\ doublet, corresponding to $-3.25 \le \log{n_H} \le -3.05~[\cc]$.  The strength of the \Lya\ line implies a metallicity of $\log{Z/\Zsun} \ge -1.5$ for the low ionization phase.  For the {\CIV} phase, low metallicities ($\log{Z/\Zsun} = -2.5$ for the blueward component and $\log{Z/\Zsun} = -1.7$ for the redward component) fit the {\Lya} profile, but higher metallicities and a separate {\Lya} phase are also permitted.

\subsection{HE0151-4326 $z=1.708494$}
\label{sec:S6}

This is a single cloud absorber in \MgII.  Due to blends at the expected locations of the \CIVdblt\ doublet and the \SiIVdblt\ doublet, it is not possible to absolutely determine if a second, high ionization phase is needed.  The ionization parameter for the low ionization phase is constrained to be $\log{U} \ge -4.0$, which is based upon the observed strength of \FeII\ absorption.  Using a number density of photons of $\log{n_{\gamma}} = -4.74$, we find $\log{n_H} \le -0.7~[\cc]$. For the high phase, if used, we find an ionization parameter of $\log{U} = -2.3$.  This corresponds to $\log{n_H} = -2.4$.  A metallicity of $\log{Z/\Zsun} \ge -1.5$ is required in order not to overproduce \Lya\ in the red wing, however, the blue wing of \Lya\ requires a separate, extremely low metallicity phase to be fit.

\subsection{HE2347-4342 $z=1.796237$}
\label{sec:S7}

This is a very weak, narrow single cloud absorber in \MgII, though a very weak second component improves the fit in the blue wing.  Due to blends at the expected locations of the \CIVdblt\ doublet and the \SiIVdblt\ doublet, there is some ambiguity in our assessment of a second phase.  However, because the \SiIV\ and \CIV\ are weak, for some parameter choices it is possible for them to arise in the same phase with the \MgII.  There are two possibilities for modeling the \MgII\ phase.  First, we can use an ionization parameter of $\log{U} \ge -3.2$ to match the observed \FeII\ but decrease the abundance of Si and Al relative to the Solar value to fit the observed \SiII\ and \AlII.  Alternatively, we can match the observed abundance of \SiIII\ by using an ionization parameter $-4.0 \le \log{U} \le -3.2$ but decrease the abundance of Si, Al, and Fe.  Using a number density of photons of $\log{n_{\gamma}} = -4.72$, we find $\log{n_H} \le -1.5~[\cc]$ or $-1.5 \le \log{n_H} \le -0.7~[\cc]$, respectively.  If a second phase is used to fit the high ionization transitions, the parameters are $-2.0 \le log{U} \le -1.0$, so as not to overproduce \SiIV.  This corresponds to $-2.7 \le \log{n_H} \le -3.7$.  The metallicity of the low ionization phase is constrained to be $\log{Z/\Zsun} \ge -1.0$.  However, the observed \Lya\ absorption  can not be fully matched without using a separate, extremely low metallicity phase, even with the addition of the broader high ionization phase.

\subsection{Q0453-423 $z=1.858380$}
\label{sec:S8}

This is a multiple cloud absorber in \MgII\ with six resolved components.  A second phase is needed to fit \SiIII\ and \SiIV\ $\lambda$1403.  Though \CIV\ is badly blended, we know that it is relatively weak, classifying this as a \CIV\ deficient system. The ionization parameter of the low ionization phase is constrained to be $\log{U} = -4.0$ in order to produce the observed \FeII\ absorption.  Using a number density of photons of $\log{n_{\gamma}} = -4.71$, we find $\log{n_H} = -0.8~[\cc]$.  This model does not produce the observed \SiIII\ or \SiIV\, and it does not produce significant \CIV\ absorption.  A second phase with $\log{U} = -2.8$ can account for the observed \SiIII\ and \SiIV\, but slightly overproduces {\MgII}, {\SiII}, and \CIV.  A slight (few tenths of a dex) abundance pattern adjustment of these elements could resolve this discrepancy. The strength of the \Lya\ line implies a metallicity of $\log{Z/\Zsun} \ge -2.0$ for the low ionization phase, but additional offset high ionization components, not constrained by these data, could contribute substantially to the \Lya\ absorption.  If so, the low ionization phase could have substantially higher metallicity.

\subsection{HE0940-1050 $z=2.174546$}
\label{sec:S9}

This is a single cloud absorber in \MgII.  A second phase is required to reproduce the observed \CIV\ absorption.  The ionization parameter is constrained to be $\log{U} \ge -2.5$ for the \MgII\ phase, and $-2.0 \le \log{U} \le -1.0$ for the five clouds in the \CIV\ phase.  The bluest cloud in the \CIV\ phase has the lowest ionization parameter.  A higher ionization parameter in the low phase will overproduce high ionization transitions, however, lower ionization parameters require a reduction in the abundance of aluminum and iron by up to 0.7 dex.  The ionization parameter in the high phase is constrained by the observed absorption of \SiIV.  Using a number density of photons of $\log{n_{\gamma}} =-4.69$, we find $\log{n_H} \le -2.2~[\cc]$ for the \MgII\ phase and $-3.7 \le \log{n_H} \le -2.7~[\cc]$ for the five clouds in the \CIV\ phase.  The strength of the \Lya\ line implies a metallicity of $\log{Z/\Zsun} \ge -2.0$.  Another phase is required to match the blue wing of the {\Lya} profile.

\section{Discussion}
\label{sec:disc}

We have compared the basic and derived properties of weak \MgII\ absorbers over the redshift range $0.4 < z < 2.4$.  We include the 9 systems at $1.4 < z < 2.4$ from the survey of \citet{Lynch06} as well as systems from \citet{Rig02,Char03,Zonak04,Ding05,Mas05} with $z < 1.4$.  We consider single and multiple cloud weak \MgII\ absorbers separately, since they are likely to have different origins.

One of the most important basic properties of single cloud weak \MgII\ absorbers at $0.4 < z < 1.4$ is
their two phase structure.  \MgII\ is found to arise in a higher density region, while the strength
of the \CIV\ absorption requires a separate, lower density region \citep{Rig02}.  This same two phase
structure is also found in all four single cloud weak \MgII\ absorbers at $1.4 < z < 2.4$ for which it
was possible to place a constraint on a second phase.  The other two single cloud weak \MgII\ absorbers
in the $1.4 < z < 2.4$ sample had blends at the expected location of \CIV\ that prevented us from
deriving a constraint.
For comparison, the weaker extragalactic background radiation (EBR) at $z\sim 0$ would lead to broader \MgII\ components arising from the high ionization phase of single cloud weak \MgII\ absorbers, and to detectable \MgII\ absorption from some structures that at higher redshift produced only high ionization absorption \citep{Nar05}.

Most $0.4 < z < 1.4$ multiple cloud weak \MgII\ absorbers also require separate phases to explain
simultaneously the observed \MgII\ and \CIV\ absorption.  Our three multiple cloud weak \MgII\
absorbers at $1.4 < z < 2.4$ also required two phase models.  At both redshift regimes, we see
examples of \CIV -deficient multiple cloud weak \MgII\ absorbers, where a second phase may
not be needed (e.g., the $z=0.5584$ system toward PG$1241+176$ \citep{Ding05} and the $z=0.7290$
system toward PG$1248+401$ \citep{Mas05}) or is needed, but produces only weak absorption
(our $z=1.796237$ system toward HE2347-4342).

Thus there may be a difference between single and multiple cloud weak \MgII\ absorbers in the
fraction that have a second, lower density phase producing relatively significant \CIV\ absorption.
Also, there seems to be a significant difference in the nature of the second phase in cases where
it is required.  For single cloud weak \MgII\ absorbers, both at $0.4 < z < 1.4$ and at $1.4 < z < 2.4$,
there is always a \CIV\ cloud centered on the \MgII\ (within $\sim 3$~{\kms}).  There may also be
additional, offset \CIV\ clouds, which tend to be weaker.  The multiple cloud weak \MgII\ absorbers
do not usually have a direct correspondence between the \CIV\ and the \MgII\ clouds.  In both types
of absorbers, we might postulate a sheetlike or shell geometry, with separate layers responsible for
the \MgII\ and \CIV\ absorption.  However, in the case of the single cloud \MgII\ absorbers it would
appear that the layers are quite quiescent, and are moving in unison.  This would
suggest an origin in an environment that has not experienced recent star formation or turbulence.



\begin{figure}
\centering
\epsscale{1.0}
\plotone{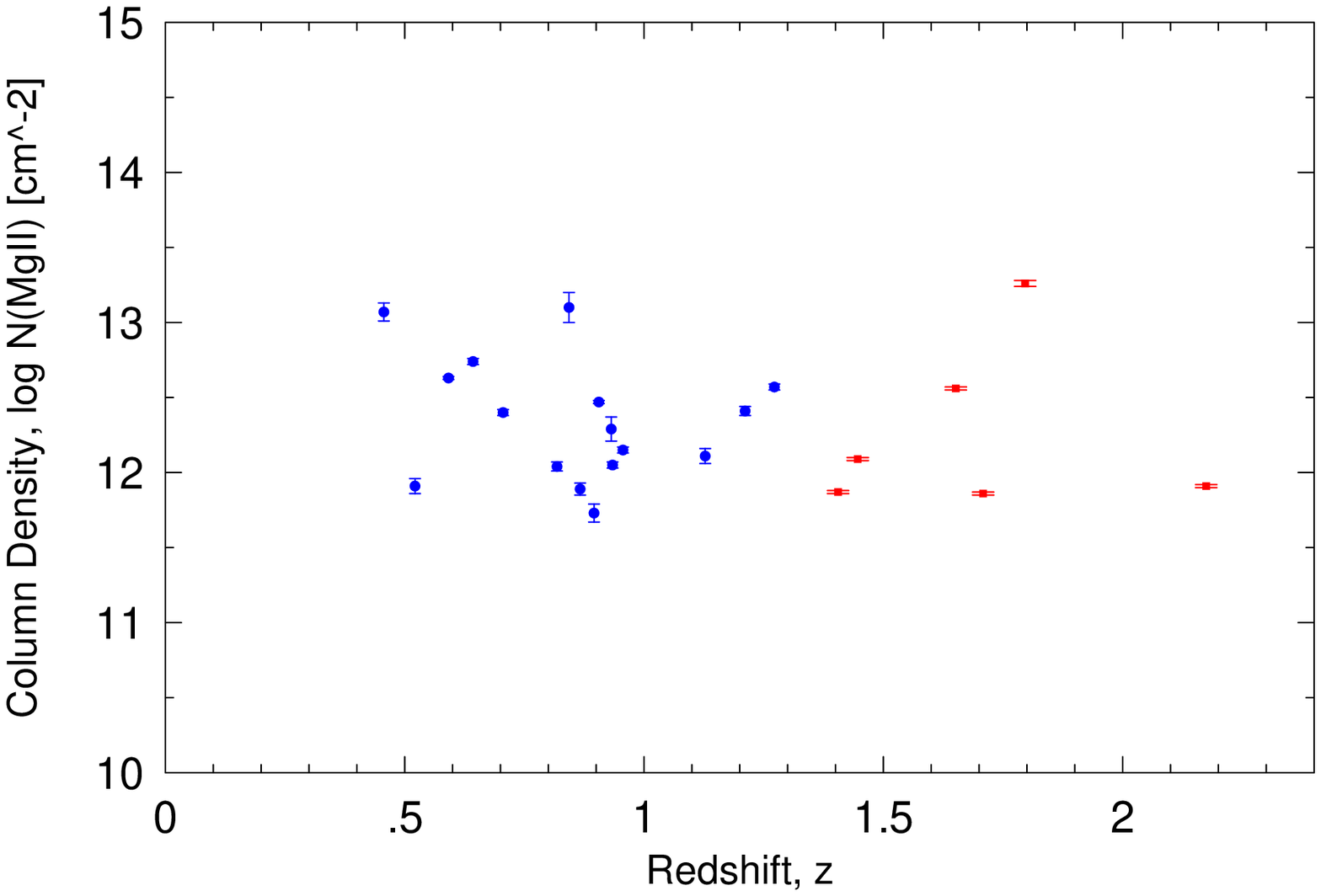}
\caption[$\log{N(\MgII)}$ vs. $z$ for single cloud weak \MgII\ absorbers]{\small Column density vs. redshift for single cloudy weak \MgII\ absorbers.  Systems below $z = 1.4$ (circles) were taken from \citet{Rig02,Zonak04,Ding05,Mas05}, while systems above $z = 1.4$ (squares) were taken from \citet{Lynch06}.  There is no statistical difference between the two populations, giving credence to the theory that the same physical processes have been responsible for the formation of weak  \MgII\ absorbers (namely star formation activity in dwarfs) across the given redshift range. \label{fig:SCNH}}
\end{figure}

\begin{figure}
\centering
\epsscale{1.0}
\plotone{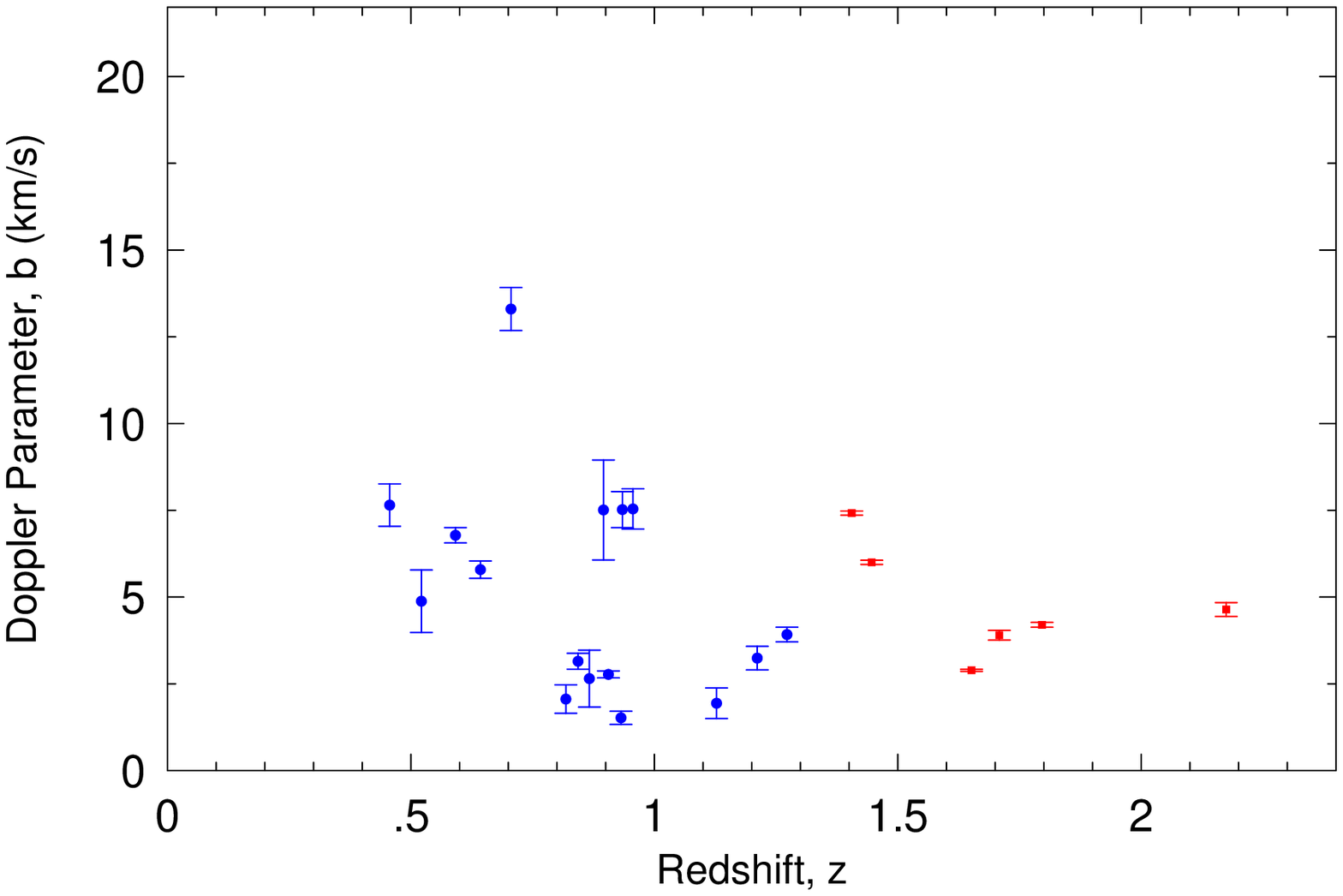}
\caption[$b$ vs. $z$ for single cloud weak \MgII\ absorbers]{\small Doppler parameter vs. redshift for single cloudy weak \MgII\ absorbers.  Systems below $z = 1.4$ (circles) were taken from \citet{Rig02,Zonak04,Ding05,Mas05}, while systems above $z = 1.4$ (squares) were taken from \citet{Lynch06}.  There is no statistical difference between the two populations, giving credence to the theory that the same physical processes have been responsible for the formation of weak  \MgII\ absorbers (namely star formation activity in dwarfs) across the given redshift range. \label{fig:SCb}}
\end{figure}

\begin{figure}
\centering
\epsscale{1.0}
\plotone{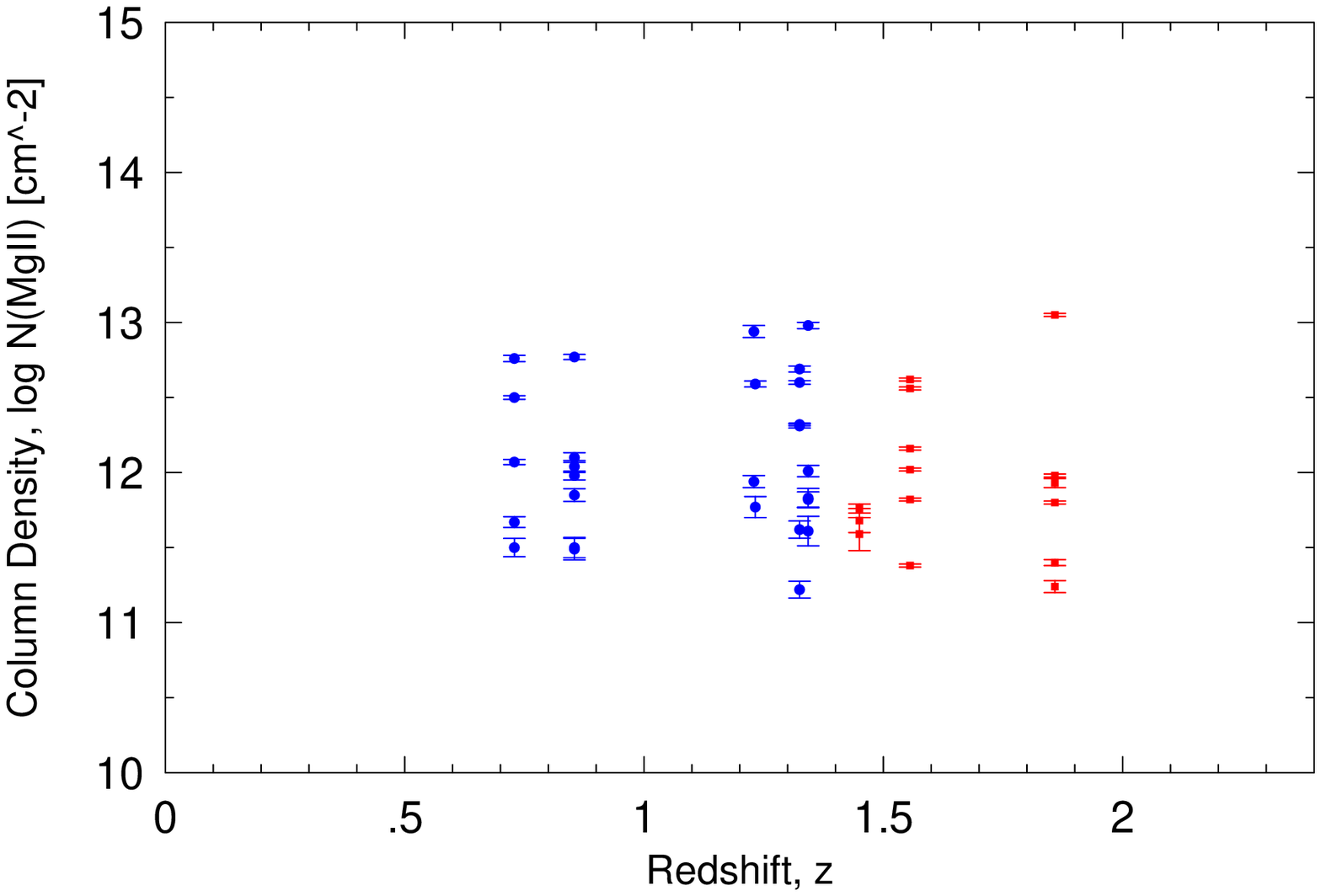}
\caption[$\log{N(\MgII)}$ vs. $z$ for multiple cloud weak \MgII\ absorbers]{\small Column density vs. redshift for multiple cloudy weak \MgII\ absorbers.  Systems below $z = 1.4$ (circles) were taken from \citet{Rig02,Zonak04,Ding05,Mas05}, while systems above $z = 1.4$ (squares) were taken from \citet{Lynch06}.  There is no statistical difference between the two populations, giving credence to the theory that the same physical processes have been responsible for the formation of weak  \MgII\ absorbers (namely star formation activity in dwarfs) across the given redshift range. \label{fig:MCNH}}
\end{figure}

\begin{figure}
\centering
\epsscale{1.0}
\plotone{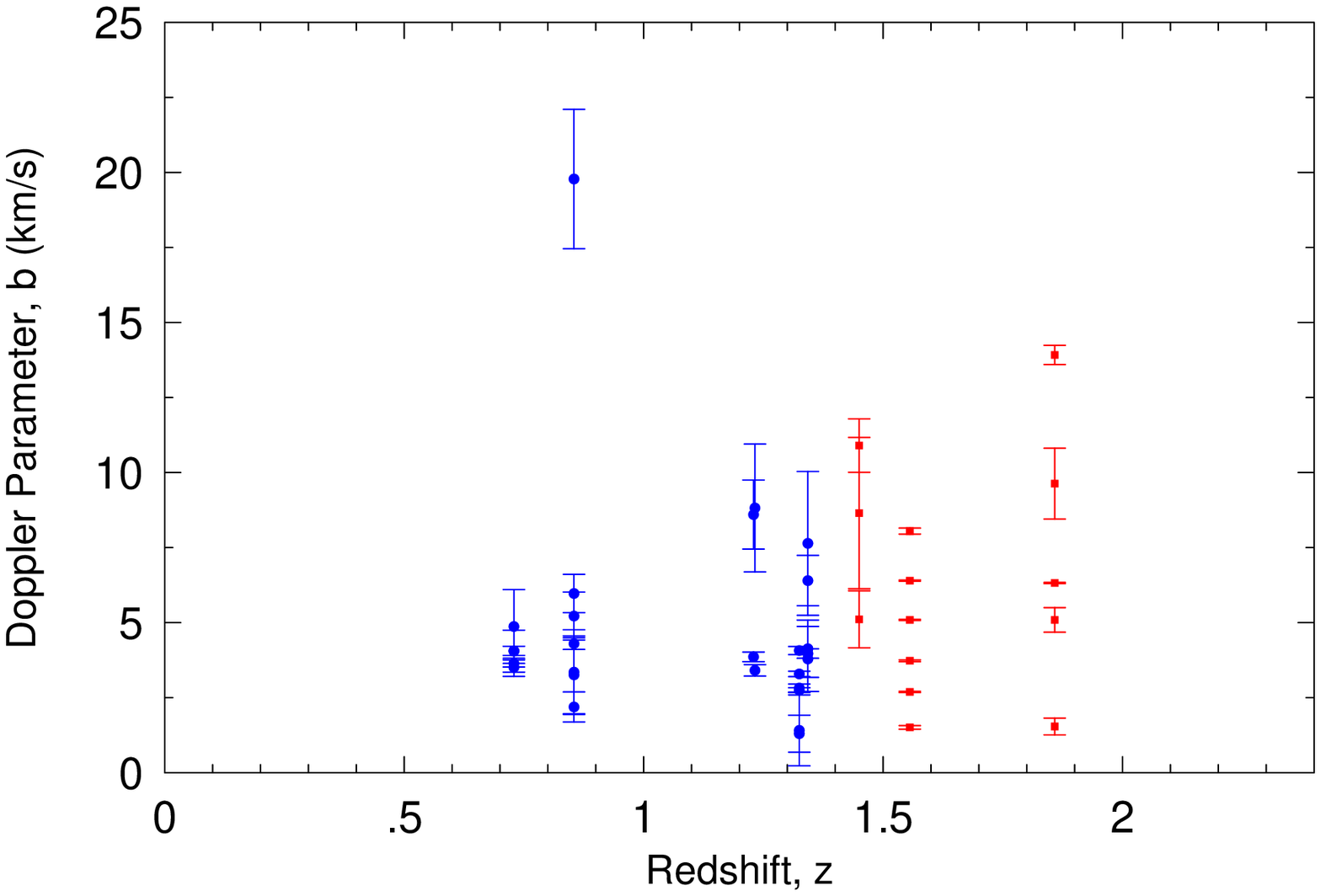}
\caption[$b$ vs. $z$ for multiple cloud weak \MgII\ absorbers]{\small Doppler parameter vs. redshift for multiple cloudy weak \MgII\ absorbers.  Systems below $z = 1.4$ (circles) were taken from \citet{Rig02,Zonak04,Ding05,Mas05}, while systems above $z = 1.4$ (squares) were taken from \citet{Lynch06}.  There is no statistical difference between the two populations, giving credence to the theory that the same physical processes have been responsible for the formation of weak  \MgII\ absorbers (namely star formation activity in dwarfs) across the given redshift range. \label{fig:MCb}}
\end{figure}

Figure \ref{fig:SCNH} shows \MgII\ column density vs. $z$ for single cloud systems, and Figure \ref{fig:SCb} shows their Doppler parameter vs. $z$.  There is no apparent change in these parameters across redshift.  The same can be said for Figures \ref{fig:MCNH} and \ref{fig:MCb}, which show column density vs. $z$ and Doppler parameter vs. $z$ for multiple cloud systems, respectively.  Table \ref{table:bandN} also gives this information. 

\begin{figure}
\centering
\epsscale{1.0}
\plotone{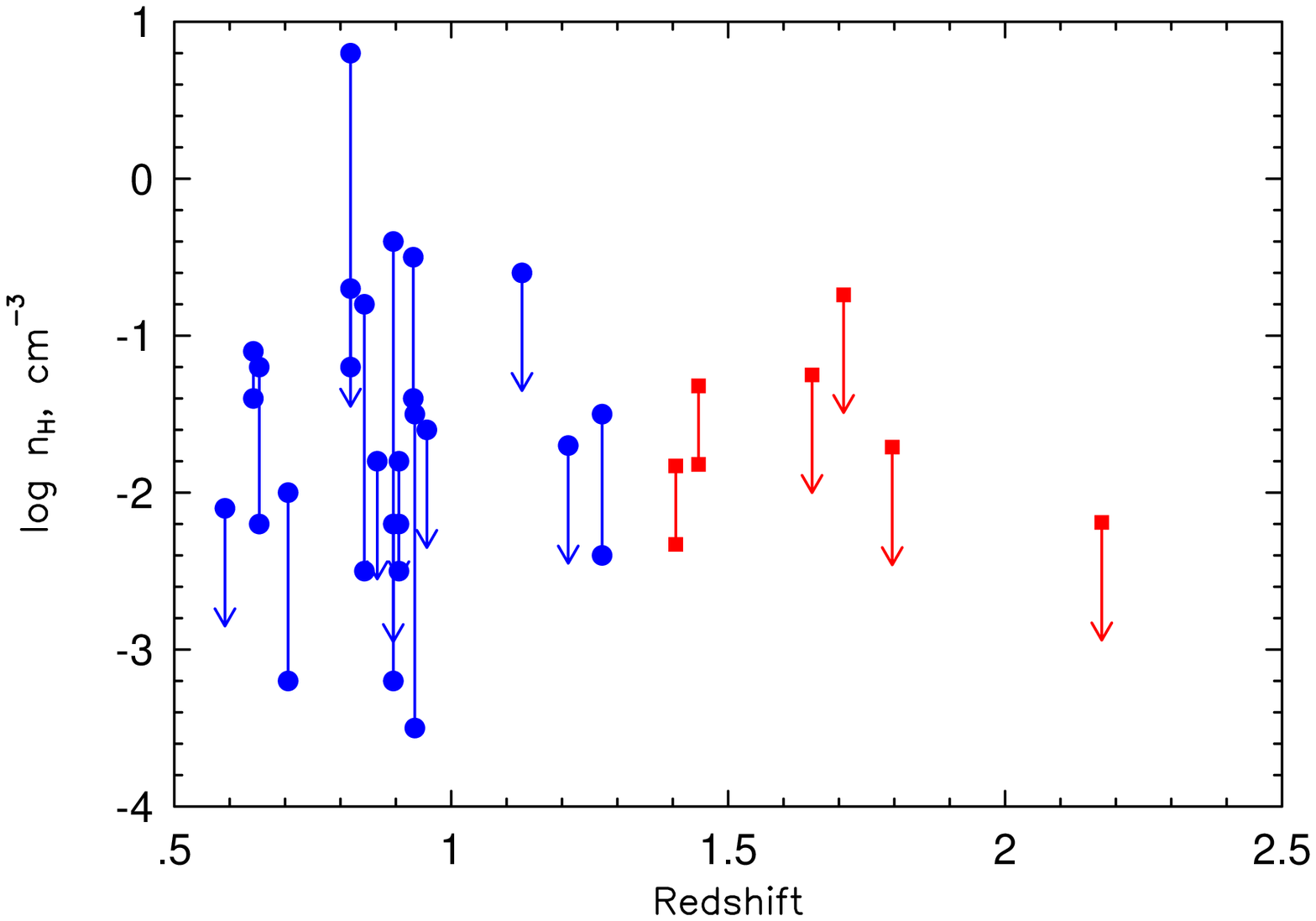}
\caption[$\log{n_H}$ vs. $z$ for single cloud weak \MgII\ absorbers]{\small Density vs. redshift for single cloudy weak \MgII\ absorbers.  Systems below $z = 1.4$ (circles) were taken from \citet{Rig02,Char03,Ding05}, while systems above $z = 1.4$ (squares) were taken from \citet{Lynch06}.  There is no statistical difference between the two populations, giving credence to the theory that the same physical processes have been responsible for the formation of weak  \MgII\ absorbers (namely star formation activity in dwarfs) across the given redshift range. \label{fig:SCnH}}
\end{figure}

Figure \ref{fig:SCnH} shows $\log{n_H}$ vs. $z$ for systems with a single absorption component in \MgII.  There is a large spread in the derived properties for the low redshift absorbers, but no systematic trend is apparent across redshift.  In most cases, only upper limits could be obtained for the density.  However, limits for $\log{n_H}$ of the high redshift systems are consistent with those of the low redshift systems. To verify that there is no significant evolution, we applied the Spearman-Kendall nonparametric rank correlation tests, which take into account the upper limits in the data \citep{isobe86,lavalley92}.  The Spearman and Kendall tests showed 62\% and 85\% chances that a correlation is not present.

\begin{figure}
\centering
\epsscale{1.0}
\plotone{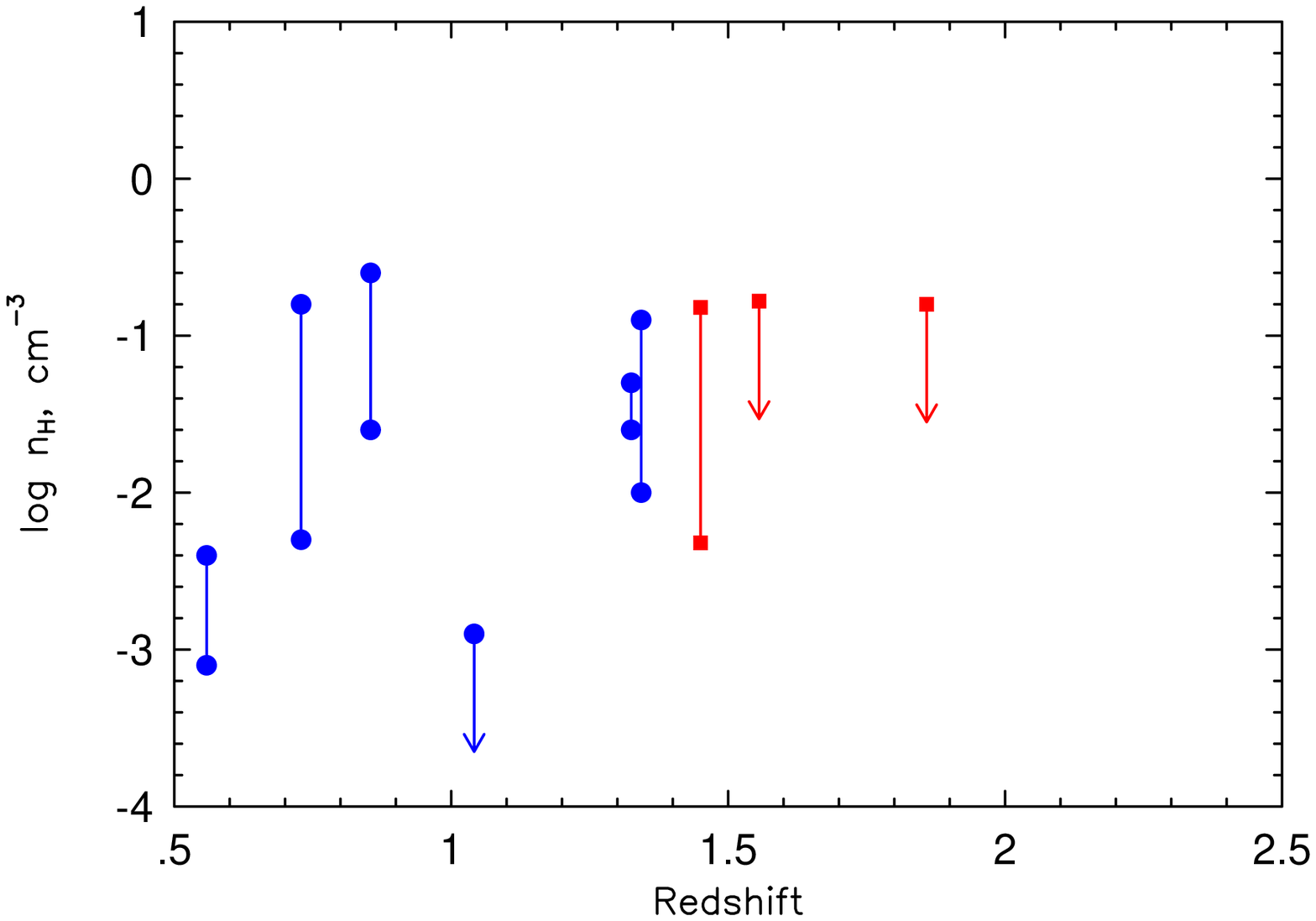}
\caption[$\log{n_H}$ vs. $z$ for multiple cloud weak \MgII\ absorbers]{\small Density vs. redshift for multiple cloudy weak \MgII\ absorbers.  Systems below $z = 1.4$ (circles) were taken from \citet{Rig02,Zonak04,Ding05,Mas05}, while systems above $z = 1.4$ (squares) were taken from \citet{Lynch06}.  There is no statistical difference between the two populations, giving credence to the theory that the same physical processes have been responsible for the formation of weak  \MgII\ absorbers (namely star formation activity in dwarfs) across the given redshift range. \label{fig:MCnH}}
\end{figure}

\begin{figure}
\centering
\epsscale{1.0}
\plotone{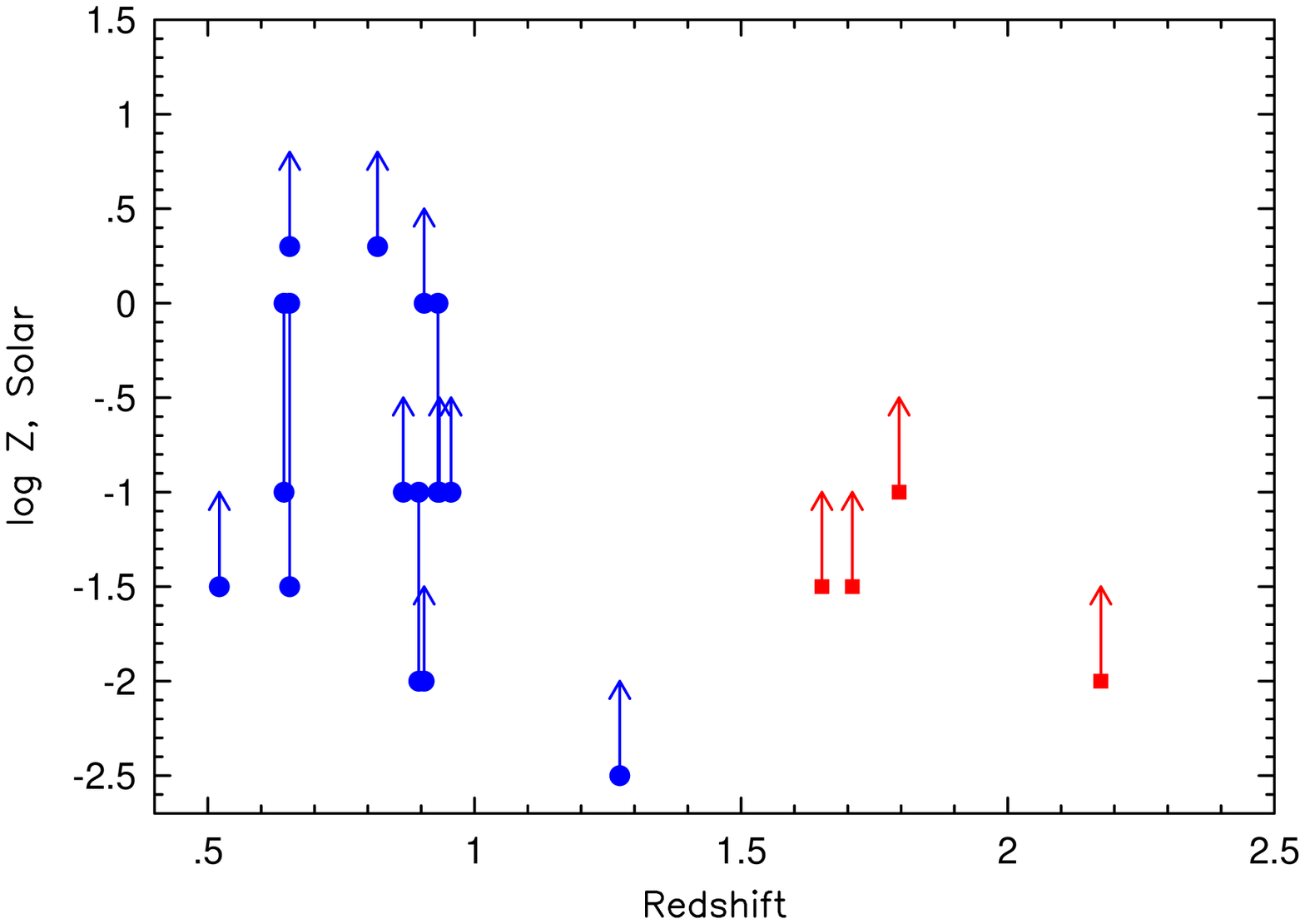}
\caption[$\log{Z/\Zsun}$ vs. $z$ for single cloud weak \MgII\ absorbers]{\small Metallicity vs. redshift for single cloudy weak \MgII\ absorbers.  Systems below $z = 1.4$ (circles) were taken from \citet{Rig02,Char03,Ding05}, while systems above $z = 1.4$ (squares) were taken from \citet{Lynch06}.  There is no statistical difference between the two populations, giving credence to the theory that the same physical processes have been responsible for the formation of weak  \MgII\ absorbers (namely star formation activity in dwarfs) across the given redshift range. \label{fig:SCZ}}
\end{figure}

\begin{figure}
\centering
\epsscale{1.0}
\plotone{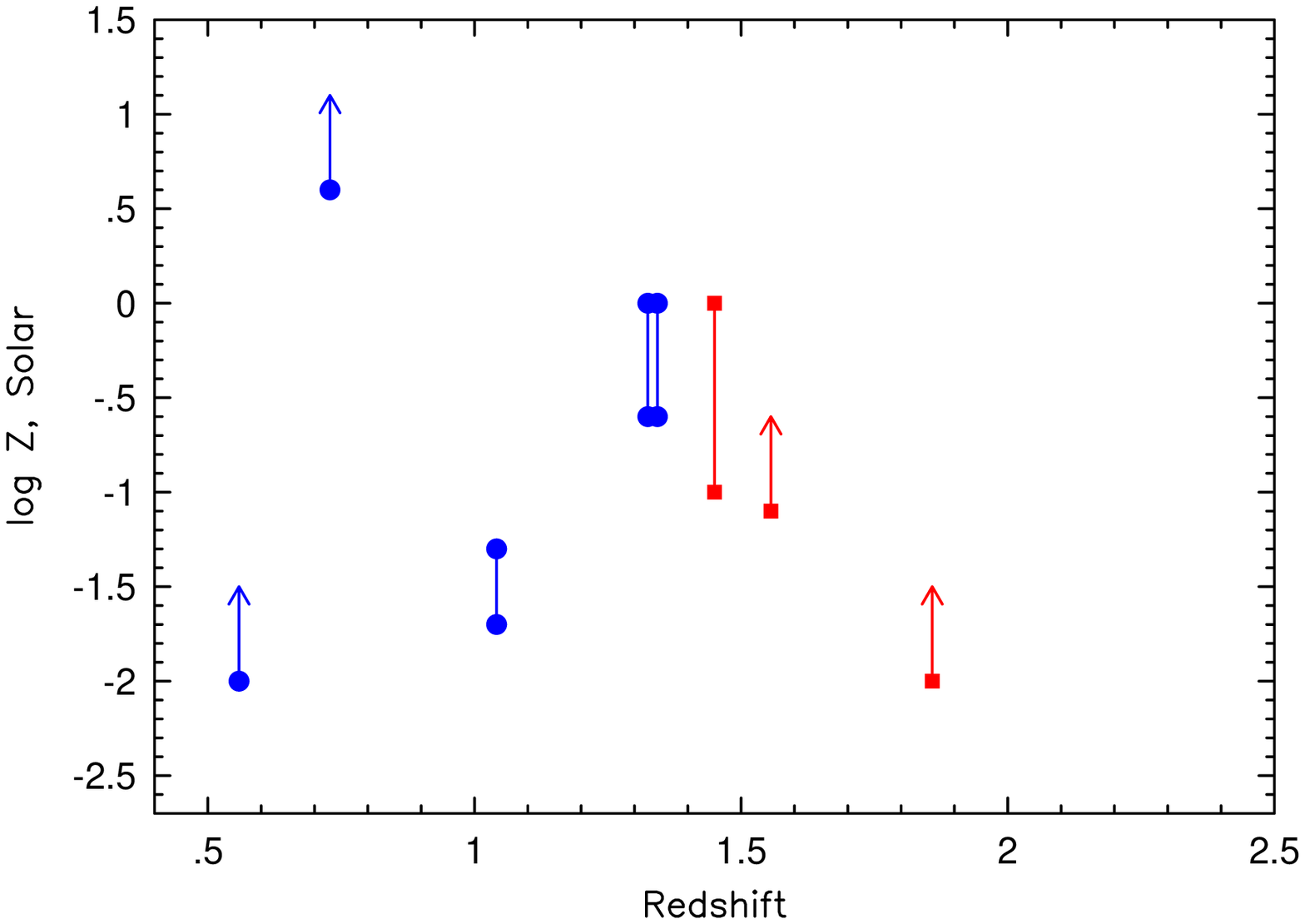}
\caption[$\log{Z/\Zsun}$ vs. $z$ for multiple cloud weak \MgII\ absorbers]{\small Metallicity vs. redshift for multiple cloudy weak \MgII\ absorbers.  Systems below $z = 1.4$ (circles) were taken from \citet{Rig02,Zonak04,Ding05,Mas05}, while systems above $z = 1.4$ (squares) were taken from \citet{Lynch06}.  There is no statistical difference between the two populations, giving credence to the theory that the same physical processes have been responsible for the formation of weak  \MgII\ absorbers (namely star formation activity in dwarfs) across the given redshift range. \label{fig:MCZ}}
\end{figure}

Figure \ref{fig:MCnH} shows $\log{n_H}$ vs. $z$ for multiple cloud absorbers.  Although our sample size is small, again there is no obvious change in the properties of the absorbers across redshift.  Figure \ref{fig:SCZ} shows $\log{Z/\Zsun}$ vs. $z$ for single cloud absorbers.  Metallicity constraints could not be obtained for all absorbers because there was not always coverage of the \Lya\ line.  Taking into account the limits, we cannot see a significant change in the properties with redshift.  The Spearman-Kendall tests yielded a large probability (0.80 and 0.33) that there is no correlation.  However, it is worth noting that we do yet know of a $z>1.4$ absorber with a high (close to solar) metallicity.  There are a few solar or higher metallicity absorbers (25\% of the sample) at $z<1.4$.  Figure \ref{fig:MCZ} shows $\log{Z/\Zsun}$ vs. $z$ for multiple cloud absorbers.  Once again, we suffer from a small sample size, but the metallicities of the high redshift systems are consistent with those of the low redshift systems.

First, we consider the possible implications of our results on the multiple-cloud absorbers.
This class can be broadly grouped into two categories.  First, there are those multiple cloud absorbers that are ``kinematically spread'' and are likely to be ``almost-strong'' \MgII\ absorbers for which the line-of-sight simply does not pass through dense regions of gas.  Second, there are those multiple cloud absorbers that are ``kinematically compact'' and are likely dwarf galaxies or are  associated with dwarf galaxies \citep{Zonak04,Ding05,Mas05}. The $z = 1.450109$ system toward Q0122-380 is an example of a kinematically compact absorber.  The metallicity of this system is constrained to be $-1.0 \le \log{Z/\Zsun} \le 0.0$.   The $z = 1.555845$ system towards HE2217-2818 and the $z = 1.858380$ system toward Q0453-423 are examples of kinematically spread absorbers.  The metallicities of these two systems are constrained to be $\log{Z/\Zsun} \ge -1.1$ and $\log{Z/\Zsun} \ge -2.0$.  Because the metallicities of our systems are not well constrained, we cannot draw any definite conclusions about the environments in which each type of system arises.

The redshift path density of single cloud weak \MgII\ absorbers is observed to decrease between $z \sim 1$ and $z \sim 2$ \citep{Church99b,Lynch06}.  Some of this evolution is due to the changing EBR which ranges from $-4.83 < \log{n_{\gamma}} < -4.71~ [\cc]$ between $1.4 < z < 2.4$, respectively.  The effect of the changing EBR is to lead to more low ionization \MgII\ gas at lower redshift.  In addition, cosmological effects will lead to a decrease in the density of weak \MgII\ absorbers at lower redshift.  When these two competing effects are taken together, they cannot fully account for the lower $dN/dz$ at $z \sim 2$.  The range of physical conditions that were found in this study (column density, Doppler parameter, density, and metallicity) for systems at redshift $1.4 < z < 2.4$ do not show a statistical variation from systems at redshift $0.4 < z < 1.4$.  The ranges are large, constraints are derived using different transitions at different redshifts, and our samples are small, leading to dilution of any trends.  However, at face-value our result is consistent with the idea that the evolution in the weak \MgII\ absorber population from $z\sim 2$ to $z\sim 1$ is due to an increase in the efficiency of the mechanisms that create weak \MgII\ absorbers, and not due to a change in the actual mechanisms.  For example, if a collapse process gave rise to weak \MgII\ absorbing structures, then, to first order, one would expect a constant range of densities across redshift.  Since we see such a constant range, the observed evolution in $dN/dz$ would then be attributed to a change in the number of structures undergoing such a collapse as a function of redshift.

We now turn our attention to the effects of a changing metallicity.  As metallicity generally increases with decreasing redshift, we would expect that, at low redshift, lower total hydrogen column density absorbers cold give rise to weak \MgII\ absorption.  This would lead to a rise in the number of weak \MgII\ absorbers.  In our data set, although there is no statistically significant trend, an increase in metallicity with decreasing redshift is still consistent with the data. Because we have a relatively small data set and only have limits in most cases, we cannot draw a firm conclusion on the change in metallicity across redshift. Thus it is possible that the increase in weak \MgII\ absorbers at lower redshift is at least in part due to a systematic increase in metallicity.  This may be a fruitful avenue for future study. To improve metallicity constraints, we would need access to lower Lyman series lines.  The narrow, low ionization components have a dominant contribution to these Lyman series lines, while the \Lya\ absorption can have contributions from broader components \citep[see Figure 4 therein]{Church99a}.

In summary, through our modeling we have found that the properties of single cloud weak \MgII\ absorbers at $1.4 < z < 2.4$ are similar to those of single cloud weak \MgII\ absorbers at $0.4 < z < 1.4$.  These properties include the existence of two phases, the gas densities, the Doppler parameters, the relatively high metallicities, and the presence of offset {\CIV} components.  It is striking that the dominant {\CIV} component, although produced in a different phase, is centered at the same velocity as the {\MgII} cloud.  
Using the facts that almost all {\CIV} absorbers are found within $\sim 100$~kpc of luminous galaxies and that half of {\CIV} absorbers have weak {\MgII} absorption, \cite{Milni05} argued that weak {\MgII} absorbers are likely to arise $\sim 50$-$100$~kpc from luminous galaxies.

Of course, the fundamental goal of our study is to identify single cloud weak {\MgII} absorbers with a specific environment and physical process.  The lack of evolution in their properties suggests a common mechanism working over time.  The various possibilities include shells or supernova remnants in dwarf galaxies, high velocity clouds, and shells of enriched material surrounding galaxies in the cosmic web.  These possibilities have in common the feature that the high and low ionization gas could be separated, but moving at the same velocity, consistent with the arguments of \citep{Milni05}.

\citet{Lynch06} note that the star formation history in dwarf galaxies seems to be consistent with the evolution of the absorber population, and suggest that this is a possible process that could give rise to weak \MgII\ absorbers.  The fact that there is no significant change in the properties of the population across redshift suggest that this scenario is possible.  We note that if this idea is correct, then star formation in the regions would have stopped long ago.  No UV photons would be left, and so we are justified in using a background spectrum rather than local stellar sources. The results of the present study are consistent with the dwarf galaxy hypothesis.

The origin of single cloud weak {\MgII} absorbers in the extragalactic analogs of high velocity clouds also remains a possibility.  The appeal of this scenario is consistency with the phase structure found in Milky Way high velocity clouds \citep{Gan05,Fox05}, the similar velocities between low and high ionization gas in the high velocity clouds, and the large covering factor of the sky by Milky Way {\OVI} high velocity clouds \citep{Sem03}.  If Milky Way high velocity clouds are produced by cool/warm clouds sweeping through the Galactic corona, a similar phenomenon would be expected to occur around other galaxies, leading to typical impact parameters of $\sim50$-$100$~kpc for lines of sight that pass through the high velocity cloud, but not through the luminous galaxy disk.  Any distinction between these sheetlike high velocity clouds structures and portions of the cosmic web clustered near galaxies may just be a matter of semantics.  Comparisons between the {\OVI} absorption in single cloud weak {\MgII} absorbers and in Milky Way high velocity clouds is a useful diagnostic, though challenging because of the location of {\OVI} in the {\Lya} forest.

\section{Conclusion}
\label{sec:conc}

We used the photoionization code Cloudy to model nine weak \MgII\ absorption systems found by \citet{Lynch06}.  The Doppler parameter and column density of \MgII\ were measured using the Minfit program \citep{Church99b}, and these were used as constraints by Cloudy.  The ionization parameter and metallicity were then adjusted incrementally in Cloudy for each \MgII\ cloud until the simulated absorption profiles matched the observed absorption profiles of other transitions in the spectra.  It was usually necessary to include a second, high ionization phase in order to reproduce the observed absorption in \CIV.  This was necessary because the \CIV\ profile was too broad and/or too strong to arise solely from the \MgII\ phase gas.  It was sometimes the case that only an upper or lower limit could be placed on the conditions of the system due to blends at the expected locations of certain transitions.  These results were then compared to models of absorbers at $0.4 < z < 1.4$ and checked for any evolution across redshift.

\begin{itemize}

\item{6/9 systems had only a single component of absorption in \MgII\ (single cloud) and the remaining three showed multiple components (multiple cloud)}
\item{A multiphase structure was required in 7/9 systems.  One system had blends at the expected location of \CIV\ and \SiIV\ and no definite conclusions about a multiphase structure could be reached.  In another system, the \CIV\ profile was weak enough that a second phase was not definitely needed, though it was preferred.}
\item{For single cloud systems we find the following constraints on physical properties:}
\subitem{For the $z = 1.405362$ absorber towards HE2347-4342, $-2.3 \le \log{n_H} \le -1.8~[\cc]$ for the low ionization phase and $\log{n_H} \le -3.1~[\cc]$ for the required high ionization phase; No metallicity constraint.}
\subitem{For the $z = 1.446496$ absorber towards Q0002-422, $-1.8 \le \log{n_H} \le -1.3~[\cc]$ for the low ionization phase and $-3.5 \le \log{n_H} \le -3.0~[\cc]$ for the required high ionization phase; No metallicity constraint.}
\subitem{For the $z = 1.651462$ absorber towards HE0001-2340, there are two possibilities for fitting the low ionization phase: $\log{n_H} \le -1.3~[\cc]$ with decreases to Si, Al, and C relative to solar, or $\log{n_H} \le -0.8~[\cc]$ with decreases to Fe, Al, and C relative to solar; $-3.3 \le \log{n_H} \le -3.1~[\cc]$ for the required high ionization phase; $\log{Z/\Zsun} \ge -1.5$.}
\subitem{For the $z = 1.708494$ absorber towards HE0151-4326, $\log{n_H} \le -0.7~[\cc]$ for the low ionization phase; Blends in high ionization transitions, but high ionization phase may not be required; $\log{Z/\Zsun} \ge -1.5$ for the low ionization phase.}
\subitem{For the $z = 1.796237$ absorber towards HE2347-4342 there are two possibilities for fitting the low ionization phase: $\log{n_H} \le -1.5~[\cc]$ or $-1.5 \le \log{n_H} \le -0.7~[\cc]$; $\log{Z/\Zsun} \ge -1.0$ for the low ionization phase; Blends in high ionization phase; weak {\CIV} could arise in same phase with \MgII, but separate phase is also permitted.}
\subitem{For the $z = 2.174546$ absorber towards HE0940-1050, $\log{n_H} \le -2.2~[\cc]$ for the low ionization phase and $-3.7 \le \log{n_H} \ge -2.7~[\cc]$ for the high ionization phase; $\log{Z/\Zsun} \ge -2.0$.}
\item{for multiple cloud systems we find the following constraints on physical properties:}
\subitem{For the $z = 1.450109$ absorber towards Q0122-380, $-2.3 \le \log{n_H} \le -0.8~[\cc]$ for the low ionization phase and $-3.2 \le \log{n_H} \le -3.0~[\cc]$ for the high ionization phase; $-1.0 \le \log{Z/\Zsun} \le 0.0$}
\subitem{For the $z = 1.555845$ absorber towards HE2217-2818, $\log{n_H} \le -0.8~[\cc]$ for the low ionization phase and $-2.8 \le \log{n_H} \le -2.3~[\cc]$ for the high ionization phase$; \log{Z/\Zsun} \ge 1.1$}
\subitem{For the $z = 1.858380$ absorber towards Q0453-423, $\log{n_H} \le -0.8~[\cc]$ for the low ionization phase and $\log{n_H} = -1.9~[\cc]$ for the high phase; $\log{Z/\Zsun} \ge -2.0$}

\end{itemize}

The properties of the absorber population as stated above are not significantly different across the redshift range $0.4 < z < 2.4$, i.e. the variation in parameters over the sample of absorbers that produce weak \MgII\ absorption is larger than any systematic evolution with redshift.  These properties include the presence of two phases to produce {\MgII} and {\CIV} absorption and the density of the gas that produces {\MgII} absorption.  With a limited number of metallicity constraints at high redshift, the data are consistent either with constant metallicity from $0.4 < z < 2.4$, or with a metallicity that increases with time.  With our increased sample size, one of the most significant results is that a required high ionization cloud is always centered within $3$~{\kms} of the single cloud weak {\MgII} absorption.

The lack of evolution in the properties of single cloud weak {\MgII} absorbers implies that the change in the number statistics of absorbers across redshift is due to changes in the rate of relevant processes and not due to a change in the nature of these processes that give rise to weak \MgII\ absorbers.  Another possibility for explaining the evolution of the number of weak \MgII\ absorbers is a systematic increase in metallicity of the absorbing structures from $z \sim 2$ to $z \sim 1$.  The close correspondence in the velocities of the low and high ionization phases suggests a layered structure which physically could be consistent with supernova remnants or winds in dwarf galaxies, or with extragalactic analogs to high velocity clouds.

\clearpage

\begin{deluxetable}{lccc}
\setlength{\tabcolsep}{0.15in}
\tablewidth{0pc}
\tablecolumns{5}
\tablecaption{Ionization Parameters and Metallicities of $1.4 < z < 2.4$ Weak \MgII\ Systems \label{table:UandZ}}
\tablehead{\colhead{QSO} & \colhead{$z_{abs}$/Velocity (\kms)} & \colhead{$\log{U}$} & \colhead{$log{Z/\Zsun}$}}
\startdata
HE2347-4342 & 1.405362 & \MgII\ Phase: -3.0 -- -2.5 & \Lya\ not covered\\
            &          & \CIV\ Phase: $\ge -1.8$ &                  \\
\hline
Q0002-422   & 1.446496 & \MgII\ Phase: -3.5 -- -3.0 & \Lya\ not covered\\
            &          & \CIV\ Phase: -1.8 -- -1.3 &                  \\
\hline
Q0122-380   & 1.450109 & \MgII\ Phase: -4.0 -- -2.5  & \Lya\ not covered\\
            &          & \CIV\ Phase: -1.8 -- -1.7  &                  \\
\hline
HE2217-2818 & 1.555845 & \MgII\ Phase: -4.0 & $\ge -1.4$\\
            &          & \CIV\ Phase: -2.5 -- -2.0 &                  \\
\hline
HE0001-2340 & 1.651462 & \MgII\ Phase: -4.0 -- $-3.5^*$ & $\ge -1.5$\\
            &          & \CIV\ Phase: -1.7 -- -1.5 &                  \\
\hline
HE0151-4326 & 1.708494 & \MgII\ Phase: $\ge -4.0$ & $\ge -1.5$\\
            &          & \CIV\ Phase: -2.3 &                  \\
\hline
HE2347-4342 & 1.796237 & \MgII\ Phase: $\ge -4.0^*$ & $\ge -1.0$\\
            &          & \CIV\ Phase: -2.0 -- -1.0 &                  \\
\hline
Q0453-423   & 1.858380 & \MgII\ Phase: -4.0 & $\ge -2.0$\\
            &          & \CIV\ Phase: -2.8 &                  \\
\hline
HE0940-1050 & 2.174546 & \MgII\ Phase: -3.7 -- -2.5 & $\ge -2.0$\\
            &          & \CIV\ Phase: -2.0 -- -1.0 &                  \\
\enddata
\tablecomments{*A more detailed description of these systems can be found in \S~\ref{sec:S5} and \S~\ref{sec:S7}}
\end{deluxetable}

\begin{deluxetable}{lccc}
\setlength{\tabcolsep}{0.15in}
\tablewidth{0pc}
\tablecolumns{5}
\tablecaption{Doppler Parameters and Column Densities of $1.4 < z < 2.4$ Weak \MgII\ Systems \label{table:bandN}}
\tablehead{\colhead{QSO} & \colhead{$z_{abs}$/Velocity (\kms)} & \colhead{$\log{N(\MgII)}~[\cmsq]$} & \colhead{$b~(\kms)$}}
\startdata
HE2347-4342 & 1.405362 & $11.87 \pm 0.01$ & $7.42 \pm 0.06$\\
\hline
Q0002-422   & 1.446496 & $12.09 \pm 0.01$ & $6.00 \pm 0.06$\\
\hline
Q0122-380   & 1.450109 &  & \\
~~Cloud 1   & -18.0 & $11.68 \pm 0.08$ & $5.11 \pm 0.95$ \\
~~Cloud 2   & -3.9 & $11.59 \pm 0.11$ & $8.65 \pm 2.52$\\
~~Cloud 3   & 40.4 & $11.76 \pm 0.03$ & $10.90 \pm 0.89$ \\
\hline
HE2217-2818 & 1.555845 &  &  \\
~~Cloud 1   & -79.7 & $11.38 \pm 0.01$ & $6.40 \pm 0.02$\\
~~Cloud 2   & -49.2 & $12.56 \pm 0.01$ & $2.69 \pm 0.02$\\
~~Cloud 3   & -30.9 & $11.82 \pm 0.01$ & $1.51 \pm 0.06$\\
~~Cloud 4   & -14.7 & $12.02 \pm 0.01$ & $8.05 \pm 0.10$\\
~~Cloud 5   & 4.9 & $12.62 \pm 0.01$ & $5.09 \pm 0.02$\\
~~Cloud 6   & 37.4 & $12.16 \pm 0.01$ & $3.73 \pm 0.03$\\
\hline
HE0001-2340 & 1.651462 & $12.56 \pm 0.01$ & $2.89 \pm 0.03$\\
\hline
HE0151-4326 & 1.708494 & $11.86 \pm 0.01$ & $3.90 \pm 0.14$\\
\hline
HE2347-4342 & 1.796237 & $13.26 \pm 0.02$ & $4.20 \pm 0.07$\\
\hline
Q0453-423   & 1.858380 & & \\
~~Cloud 1   & -26.3 & $11.24 \pm 0.04$ & $9.63 \pm 1.18$\\
~~Cloud 2   & -1.3 & $13.05 \pm 0.01$ & $6.32 \pm 0.02$\\
~~Cloud 3   & 23.1 & $11.40 \pm 0.02$ & $5.09 \pm 0.41$\\
~~Cloud 4   & 50.6 & $11.98 \pm 0.01$ & $13.92 \pm 0.32$\\
~~Cloud 5   & 100.8 & $11.80 \pm 0.01$ & $1.54 \pm 0.28$\\
~~Cloud 6   & 125.1 & $11.93 \pm 0.03$ & $33.83 \pm 2.42^*$ \\
\hline
HE0940-1050 & 2.174546 & $11.91 \pm 0.01$ & $4.64 \pm 0.20$\\
\enddata
\tablecomments{*A defect in the spectrum, most likely a sky line that was not properly removed, is responsible for this anomalous measurement}
\end{deluxetable}

\clearpage

\clearpage

\end{document}